\documentclass[preprint2,natbib209]{aastex}
\usepackage{graphicx}
\usepackage{amsmath}
\usepackage{amssymb}
\usepackage{natbib}
\usepackage{cases}
\usepackage{paralist}
\usepackage{bm}
\usepackage{upgreek}
\usepackage{bbold}
\newcommand{\taurex}{$\mathcal{T}$-REx}
\newcommand{\exomol}{{\it ExoMol}}
\newcommand{\hitran}{{\it HITRAN}}
\newcommand{\hitemp}{{\it HITEMP}}
\newcommand{\multinest}{{\ttfamily MultiNest}}
\newcommand{\occam}{{\ttfamily Occam}}
\newcommand{\marple}{{\ttfamily Marple}}

\DeclareMathAlphabet{\mathpzc}{OT1}{pzc}{m}{it}

\shorttitle{TauREx}
\shortauthors{Waldmann et al.}

\begin{document}

%@OverheardOnAph rocks!

%\title{On the blind wavelet beaten shores of sparsity: Wavelet-ICA and Exoplanets}
%\title{Exoplanet transits: sparse, redundant and blind}
\title{Tau-REx I: A next generation retrieval code for exoplanetary atmospheres }

%\author{I. P. Waldmann, G. Tinetti, E. Barton, M. Tessenyi, R. Varley, M. Hollis, S. Yurchenko, J. Tennyson }
\author{I. P. Waldmann, G. Tinetti, M. Rocchetto, E. J. Barton, S. N. Yurchenko, J. Tennyson}
\affil{Department of Physics \& Astronomy, University College London, Gower Street, WC1E 6BT, UK}
\email{ingo@star.ucl.ac.uk}

\begin{abstract}

  Spectroscopy of exoplanetary atmospheres has become a well
  established method for the characterisation of extrasolar planets. We
  here present a novel inverse retrieval code for exoplanetary
  atmospheres. \taurex~ (Tau Retrieval for Exoplanets) is a
  line-by-line radiative transfer fully Bayesian retrieval framework.
   \taurex~includes the following features: 1) the optimised use of molecular line-lists
  from the \exomol~project; 2) an unbiased atmospheric composition
  prior selection, through custom built pattern recognition software;
  3) the use of two independent algorithms to fully sample the
  Bayesian likelihood space: nested sampling as well as a more
  classical Markov Chain Monte Carlo approach; 4) iterative Bayesian
  parameter and model selection using the full Bayesian Evidence as
  well as the Savage-Dickey Ratio for nested models, and 5) the
  ability to fully map very large parameter spaces through optimal
  code parallelisation and scalability to cluster computing.  In this
  publication we outline the \taurex~framework and demonstrate, using
  a theoretical hot-Jupiter transmission spectrum, the parameter
  retrieval and model selection.  We investigate the impact of
  Signal-to-Noise and spectral resolution on the retrievability of
  individual model parameters, both in terms of error bars on the
  temperature and molecular mixing ratios as well as its effect on the
  model's global Bayesian evidence.

\end{abstract}


\section{Introduction}

Remote sensing of atmospheres and inverse retrieval methods have a well established and long standing history. Beginning with pioneering work on our own Earth \citep[e.g.][]{1969Sci...165.1256W,Conrath:1970we}, we quickly extended our grasp to other planets in our solar system \citep[e.g.][]{1972Sci...175..305H,1973JGR....78.4267C,1976RvGSP..14..609R,1981Sci...212..192H}. With the first detection of exoplanetary atmospheres \citep{Charbonneau:2002er} we have taken this work beyond our solar system confines. 

In recent years, the field of extrasolar spectroscopy has seen a increased effort in the development of data analysis and de-trending techniques \citep[e.g.][]{swain08,carter09,burke10,snellen10b,2010A&A...523A..35T,swain10,waldmann12,2013ApJ...766....7W,2012ApJ...747...12W,2014ApJ...780...23W,2012MNRAS.419.2683G,2012ApJ...761....7C, berta12,2014ApJ...786...22M,2014ApJ...785...35D,2014Natur.505...69K}. With the maturation of these methodologies, we are obtaining a rapidly increasing number of exoplanetary emission and transmission spectra requiring interpretation. We refer the reader to  \citet{2011ConPh..52..602T} and \citet{2013A&ARv..21...63T} and references within, for reviews of current spectroscopic results. 

This ever increasing wealth of spectroscopic data of extrasolar planet atmospheres allows an unprecedented insight into the properties of these foreign worlds.

The interpretation of atmospheric spectra of extrasolar planets through {\it inverse atmospheric retrieval} modelling \citep[e.g.][]{2007Icar..189..457F,Terrile:2008iv,2008JQSRT.109.1136I,2009ApJ...707...24M,Lee:2011gl,2012ApJ...749...93L,2012ApJ...753..100B,2013MNRAS.434.2616B,2014RSPTA.37230086G} has become the industry standard. 
\citet{2013ApJ...775..137L} provides a recent and comprehensive review of currently existing exoplanetary atmospheric retrieval codes.

With greater accuracy in data often comes an increased complexity in its interpretation. In analogy to recent challenges in observational exoplanetary data analysis, one can identify three major objectives for the interpretation of exoplanetary spectra: 

\underline{Sensitivity:} Given the often low resolution and low signal-to-noise (S/N) of currently available exoplanetary spectra, an understanding of the limitations and degeneracies of spectroscopic models is paramount. 

\underline{Objectivity:} Are the results driven by model dependencies, over-constraint inputs or human biases? An idealised atmospheric retrieval should make no prior assumptions about the complex nature of exoplanetary atmospheres. Whilst this is often infeasible, modern retrieval algorithms should be designed to take into account the broadest possible range of atmospheric models. It should then select amongst these models using a consistent and quantifiable metric of parameter and model adequacies. 

\underline{Big data:} With the increasing automation of exoplanet observations, the manual interpretation of atmospheric spectra will become infeasible. A modern retrieval algorithm should bear this in mind and allow for a high degree of intelligent automation and scalability to larger cluster computing.  

In this paper, we introduce a new atmospheric retrieval code, \taurex~(Tau Retrieval for Exoplanets), which has been designed with the above objectives in mind. 
Here we will describe the overall architecture and atmospheric retrieval for transmission spectroscopy and dedicate a subsequent publication (Waldmann et al. in prep.) to the emission/reflection spectroscopy case and the parameterisation of the temperature-pressure (T-P) profile.

\subsection{\taurex}

\taurex~is a novel, fully Bayesian, retrieval code for exoplanetary atmospheres. In its current implementation, \taurex~includes the following features:

\underline{Line-by-line:} \taurex~uses customised molecular and atomic
line lists available directly from the \exomol\footnote{http://www.exomol.com} project \citep{Tennyson:2012ca}. In particular, ExoMol provides computed
line lists valid over extended temperature ranges for a variety of
molecules including water \citep{BT2}, ammonia \citep{BYTe}, methane \cite{Yurchenko:2014vg} and a variety of diatomic molecules
\citep{exo1,Barton:2013eh,2014MNRAS.437.1828B,2014arXiv1403.7952B}. 
Besides line lists \exomol~ provides cross sections \citep{Hill:2012ue} for
\exomol~ and other line lists. In this work cross sections for CO, NO and
CO$_2$ created from \hitemp\ \citep{2010JQSRT.111.2139R} and TiO from
Schwenke \citep{98Scxxxx.TiO} are also used.
Molecular (and atomic) absorption cross-sections are calculated on an
optimal linear or non-linear wavelength grid resulting in a optimally
sparse cross-section library with fine griding of optically thick
lines. This guarantees high computational efficiency without loss of
accuracy.

\underline{Non-parametric prior constraint:} The priors to the
Bayesian retrieval such as number and type of molecules considered,
abundance and temperature ranges are not manually set by individual
users but automatically determined by \taurex~based on the probability
of individual molecules being present in the exoplanetary spectrum.
The \marple~module is a custom built pattern recognition package
capable of rapidly identifying likely absorbers/emitters in the
exoplanetary spectra from large line-list archives (e.g. \exomol~ \citep{Tennyson:2012ca},
\hitran\footnote{http://www.cfa.harvard.edu/hitran/} \citep{2009JQSRT.110..533R,HITRAN2012} and \hitemp\ \citep{2010JQSRT.111.2139R}).
Such an approach is highly efficient as a very large number of
molecules/atoms/ions can be considered and minimises the human bias in
selecting `key atmospheric components'.

\underline{Bayesian Model Selection:} The code can be run to integrate over the full likelihood space of the Bayesian argument allowing for the Bayesian partition function, also called the Bayesian Evidence, to be calculated. This allows for the posterior distributions of model parameters to be calculated, as well as the adequacy of the model itself, given the data, to be assessed and iteratively optimised.  

\underline{Scaleability:} Even the simplest retrieval cases can feature a high dimensional likelihood space. By relying on nested sampling approaches, we can naturally achieve an excellent multi-core processor scalability and full parallelisation of the code, allowing us to fully map possible correlations in the likelihood space.  

\section{Code overview}

The retrieval code discussed here is based on a fully modular, object oriented architecture. Amongst others, one of the main advantages of a modular approach is a more flexible and structured approach to complex programs. Throughout this paper we will follow this modular approach in describing \taurex's individual components.  

\begin{figure}[h]
\centering
\includegraphics[width=10cm]{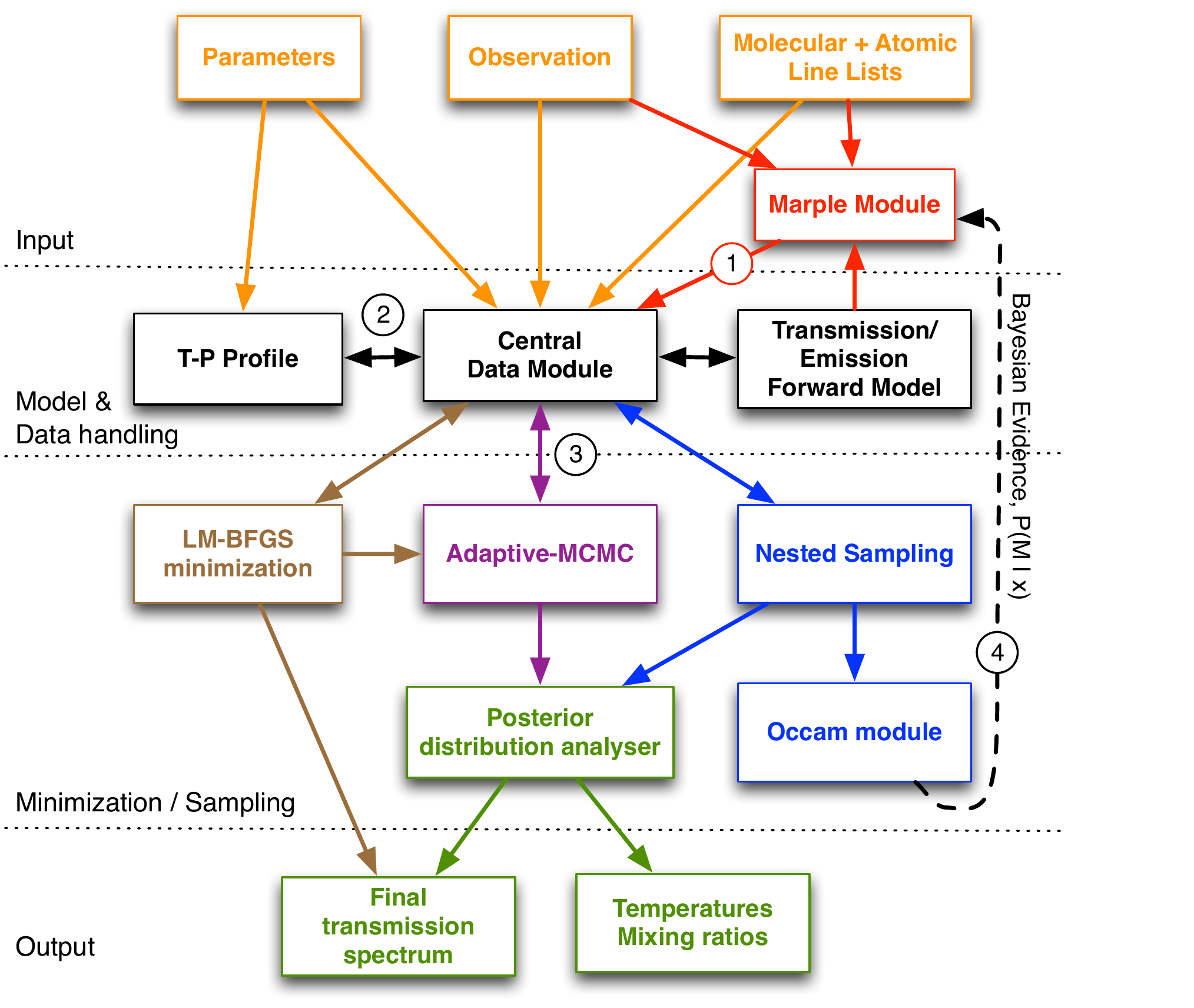}
\caption{Flowchart illustrating the modular design of \taurex. As described in the text, \taurex~is subdived into four main parts: Input, Model and data handling, Retrieval/minimization and Output analysis.    \label{fig:flowchart}}
\end{figure}

The \taurex~design is illustrated in figure~\ref{fig:flowchart} and can be broadly subdivided into four main programatic segments: 

\begin{enumerate}
\item {\it Inputs} (sections~\ref{sec:inputs}~\&~\ref{sec:marple}) - these include global parameters, the observed exoplanetary spectrum and the absorption cross-sections for the range of molecules to be considered. The \marple~module (point 1 in figure~\ref{fig:flowchart}) acts as extra input to the main code, providing a best initial guess of the atmospheric composition of the extrasolar planet to \taurex. 

\item {\it Model  and Data handling}  - defines the radiative transfer forward model, the temperature-pressure (TP) profile, overall input data handling. The Central Data Module (point 2 in figure~\ref{fig:flowchart}) acts as an abstraction layer between model, minimisation and data, allowing for an easy interchangeability of models, minimisation/retrieval techniques and data types. 

\item {\it Retrieval} (sections~\ref{sec:retrieval}) - contains three minimisation codes: 1) Limited-memory Broyden-Fletcher-Goldfarb-Shannon (LM-BFGS) algorithm, 2) Adaptive Markov Chain Monte Carlo (MCMC) sampling, 3) Bayesian Nested Sampling (NS). For NS runs, the Bayesian Partition function is calculated and model selection is performed. Point 3 in figure~\ref{fig:flowchart} illustrates that all these algorithms have a common standardised interface with the Central Data Module. This guarantees exact and comparable results for different retrieval techniques. 

The \occam~module (section~\ref{sec:occam}) performs Bayesian model selection on the outputs of the MCMC and NS. In the case of under or over-complete models, the \occam~module updates the planetary transmission model in an iterative manner (point 4 in figure~\ref{fig:flowchart}).

\item {\it Output} (section~\ref{sec:output}) - the final exoplanetary spectrum is returned along with all parameter posterior distributions, cross-correlations and Bayesian Evidence. 
\end{enumerate}

%\subsection{Code Optimization and MPI support}

%\section{Inputs}
%\label{sec:inputs}
%
%As shown in figure~\ref{fig:flowchart}, there are three main inputs to \taurex: 1) a set of user specified parameters, 2) the observed spectrum and 3) molecular/atomic line lists. 

\section{Atomic and Molecular Line-lists}
\label{sec:linelists}
\label{sec:inputs}

The optimal treatment of atomic and molecular line-lists is key to the accuracy achieved by \taurex. Throughout the code we perform line-by-line radiative transfer calculations at typically 50-100 times higher spectral resolution than the resolution of the observed spectrum to be analysed. These `high-resolution' spectra are binned down (at each iteration of the code) to the data to calculate the $\chi^{2}$ of the model fit. This ensures a correct treatment of optically thick absorption regions. In future versions of the code, we plan to include optimal non-linear binning of line-lists to further increase the computational efficiency without impacting model accuracies (Barton et al. in prep). 

\taurex~allows for an easy and seamless inclusion of large numbers of line lists. For the scope of this paper we limit ourselves to line-lists from absorption cross-sections obtained from  \exomol~ but \hitran~line-lists (or a combination of both) are equally natively supported. Automated pre-processing steps allow for conversions to a uniform data format with cross sections typically at $\Delta \nu = 1.0~ \text{cm}^{-1}$ resolution. The cross-section library is generated at temperature intervals of 100K \citep{Hill:2012ue} and upon execution of the main code interpolated to a user-set temperature resolution (typically 10K). Two forms of cross section interpolation are available: 1) linear and 2) optimal. For the optimal case, we follow \citet{Hill:2012ue} where the temperature interpolated cross-section $\varsigma_{m,\lambda}(T)$ is given by

\begin{equation}
\varsigma_{m,\lambda}(T) = a_{m,\lambda} e^{-b_{m,\lambda}/T}
\label{equ:crosssec-interp}
\end{equation}

\noindent where $m$ is the molecular/atomic species index, $\lambda$ the wavelength, $T$ the final temperature and $a$ and $b$ are scaling factors given by

\begin{equation}
b_{m,\lambda} = \left ( \frac{1}{T_{2}} - \frac{1}{T_{1}} \right )^{-1} \text{ln}\frac{\varsigma_{m,\lambda}(T_{1})}{\varsigma_{m,\lambda}(T_{2})} 
\label{equ:crosssec-interp2}
\end{equation}

\begin{equation}
a_{m,\lambda} = \varsigma_{m,\lambda}(T_{1})e^{b_{m,\lambda}/T_{1}}
\label{equ:crosssec-interp3}
\end{equation}

\noindent where $T_{1}$ and $T_{2}$ are upper and lower temperatures respectively.

\section{Forward Model}
\label{sec:forward}

The transmission forward model is based on the {\ttfamily Tau} code by \citet{2013CoPhC.184.2351H} but was optimised for a significantly higher computational efficiency. We will only give a brief summary of the transmission model and refer the interested reader to the relevant literature \citep[e.g.][]{brown01,Liou:2002uh,tinetti11c,2013CoPhC.184.2351H}. As previously mentioned, in this paper we will only describe the transmission part of \taurex~ and an isothermal temperature-pressure (T-P) profile. We dedicate a second publication (Waldmann et al. in prep.) to a complete treatment of the emission case and T-P profile parametrisation. 

The monochromatic intensity, $I_{\lambda}(z)$, of radiation passing through a gas is given by the {\it Beer-Bouguer-Lambert Law} as function of atmospheric altitude, $z$,
\begin{equation}
\label{equ:intensity}
I_{\lambda}(z) = I_{\lambda}(0)e^{-{\tau}_{\lambda}(z)}
\end{equation}

\noindent where $\lambda$ is the wavelength of the radiation, $I_{\lambda}(0)$ the incident radiation intensity at the top of the atmosphere and $\tau_{\lambda}(z)$ the optical depth of the medium. For a given absorber, $m$, we can state the optical depth to be the integral of the absorption cross-section, $\varsigma_{m}(\lambda)$, the column density, $\chi_{m}(z)$ and the number density, $\rho_{N}(z)$, over the optical path length $l(z)$

\begin{equation}
{\tau}_{\lambda,m}(z) = 2 \int_{0}^{l(z)} \varsigma_{m}(\lambda)\chi_{m}(z) \rho_{N}(z)\text{d}l
\end{equation}

\noindent where the path length is dependent on the geometry of the transmission through the planet's terminator  \citep[see figure~2 in][]{2013CoPhC.184.2351H}. The overall optical depth is now given by the sum of the individual optical depths

\begin{equation}
{\tau}_{\lambda}(z) = \sum_{m=1}^{N_{m}} {\tau}_{\lambda,m}(z)
\end{equation}

\noindent where $N_{m}$ is the total number of absorbing species, $m$. We can now calculate the equivalent atmospheric depth, $\alpha_{\lambda}$, by summing over all atmospheric depth layers, z,  

\begin{equation}
\label{equ:atmabsorb}
{\alpha}_{\lambda} = 2 \int_{0}^{z_{max}} (R_{p} + z)(1- e^{-{\tau_{\lambda}}(z)})\text{d}z
\end{equation}

\noindent where $z_{max}$ is the maximum depth of the atmosphere considered. The total transit depth as a function of $\lambda$ is hence given by 

\begin{equation}
\label{equ:transdepth}
{\delta_{\lambda}} = \frac{R_{p}^{2} + {\alpha_{\lambda}}}{R_{\ast}^{2}}
\end{equation}

\noindent where $R_{p}$ and $R_{\ast}$ are the radii of the planet and star respectively. 

\taurex~provides a full implementation of Rayleigh and Mie scattered as well as cloudy atmospheres. We refer the reader to \citet{2013CoPhC.184.2351H} for details of implementation. Retrieval degeneracies due to cloud models will be discussed in a separate publication (Rocchetto et al., in prep.).

\section{\marple~module}
\label{sec:marple}

\begin{figure}[h]
\centering
\includegraphics[width=\columnwidth]{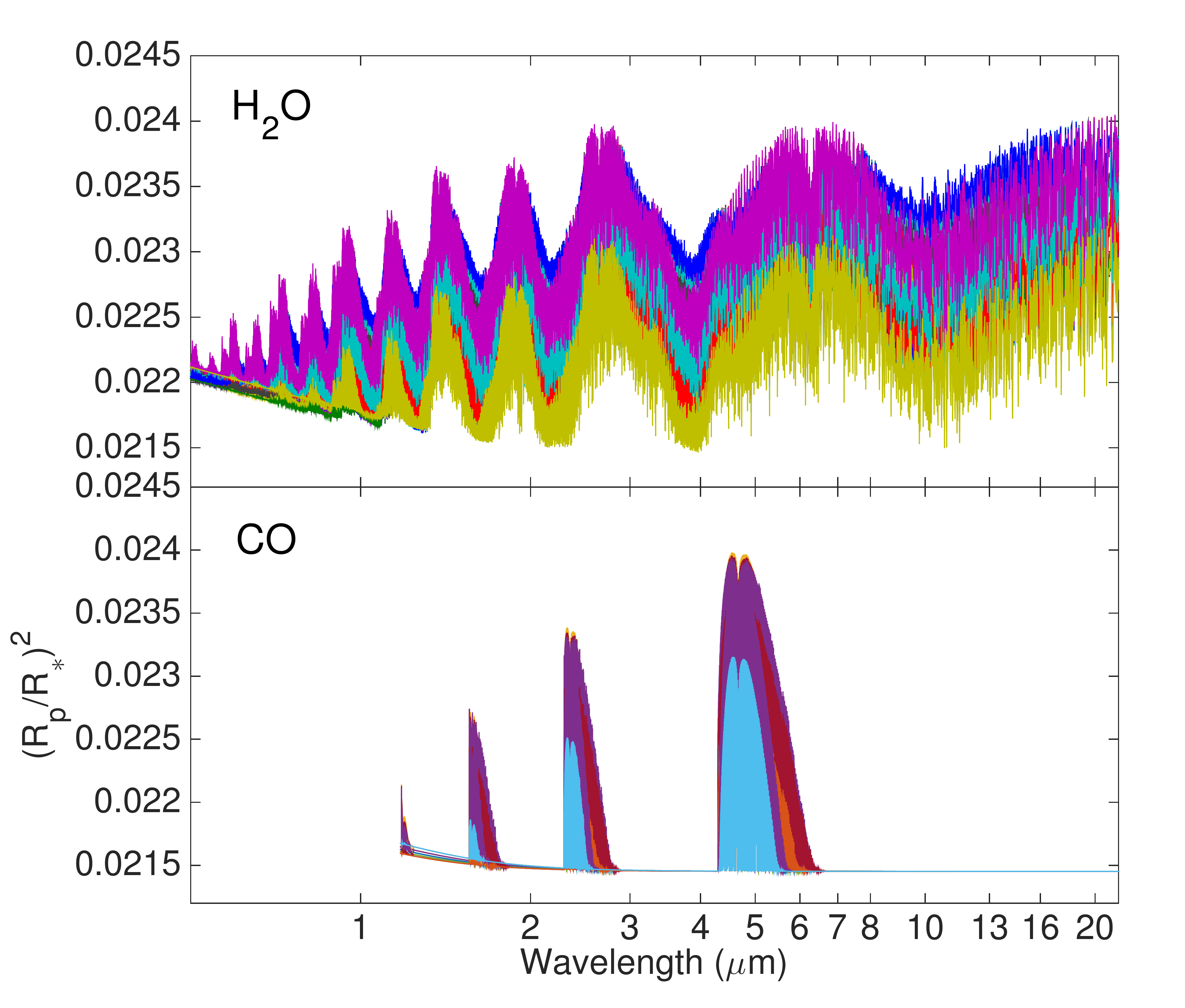}
\caption{Transmission spectra of H$_{2}$O (top) and CO (bottom) for temperature and abundance ranges of 600-2000K and 1$\times$10$^{-5}$ - 1$\times$10$^{-2}$ respectively. Planet/star and orbital parameters are taken to be similar to hot-Jupiter HD~209458b. \label{fig:water+nospectra}}
\end{figure}

\begin{figure}[h]
\centering
\includegraphics[width=\columnwidth]{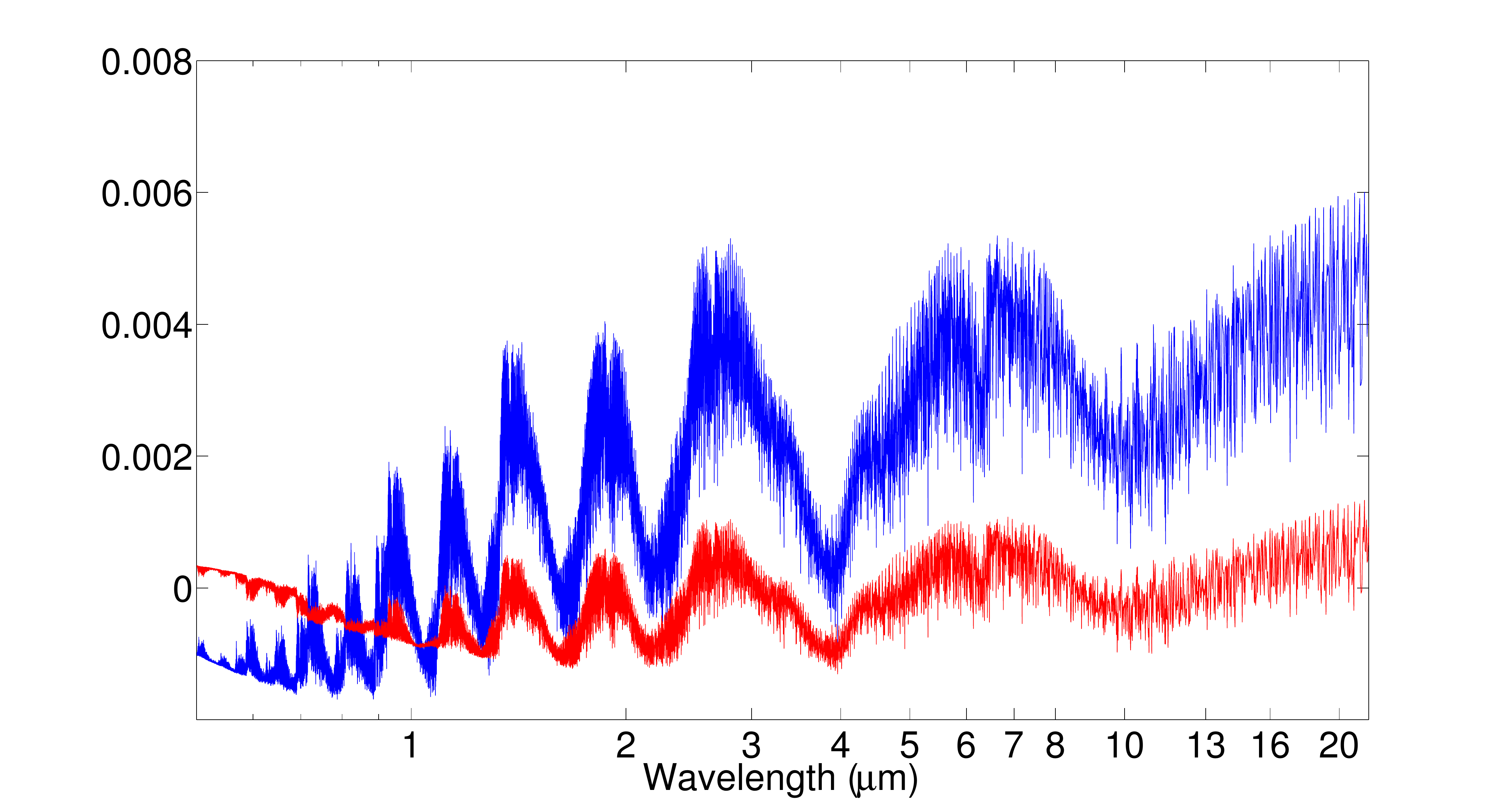}
\caption{First (blue) and second (red) principal components of the water transmission spectrum library shown in figure~\ref{fig:water+nospectra}. The first component is used to create the `feature mask', see text, whereas the second component is correlated against the observed data to identify possible matches between the observed spectra and water features.  \label{fig:waterprincipal}}
\end{figure}

\begin{figure}[h]
\centering
\includegraphics[width=\columnwidth]{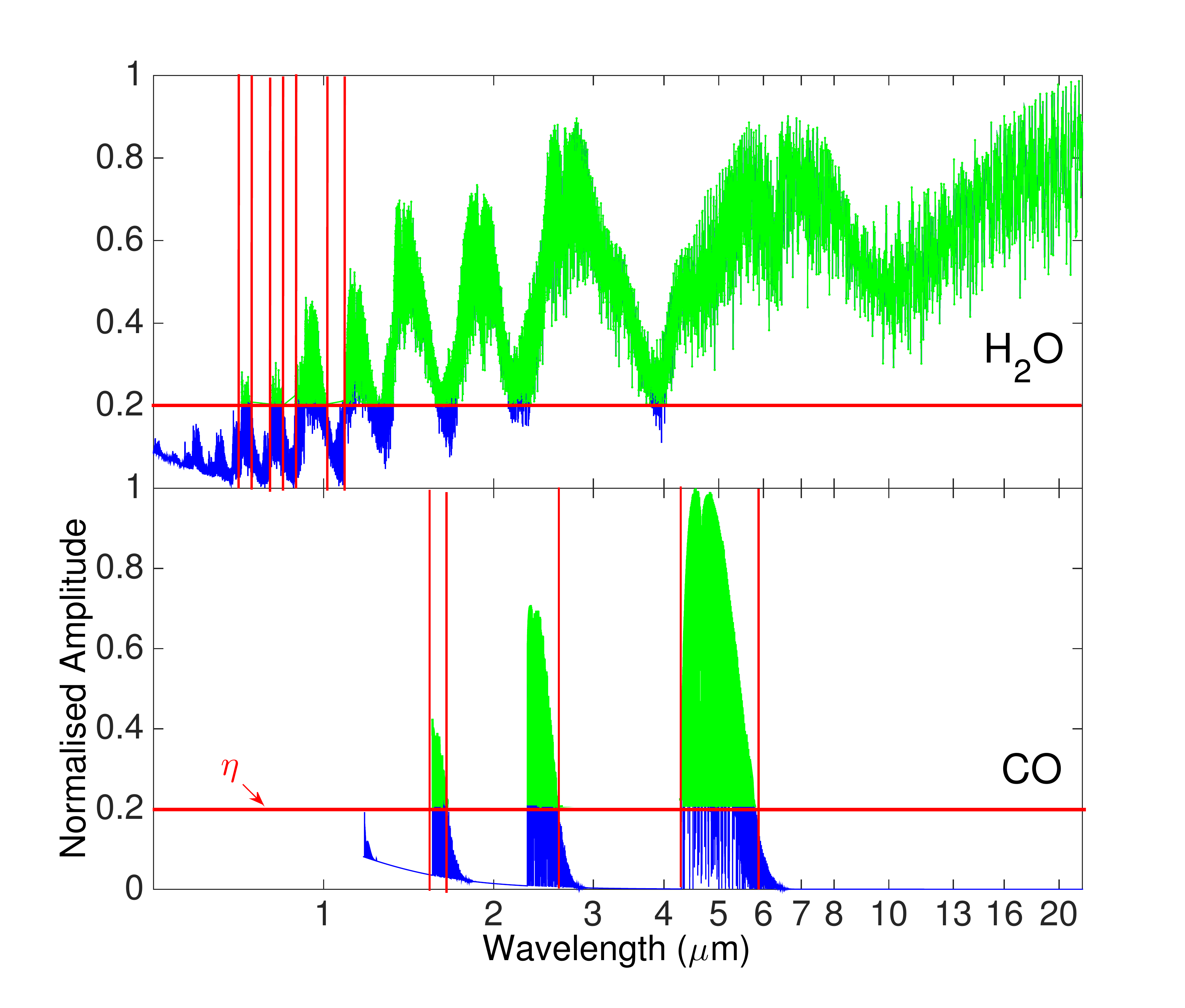}
\caption{showing the creation of the `feature mask', ${\bm \psi}_{m}(\lambda)$, for H$_{2}$O (top) and CO (bottom). In blue are the first principal components of the molecules whilst in green are the spectral features selected for the mask. The horizontal, red line indicates the threshold parameter $\eta$, equation~\ref{equ:pre-boolean}. Spectral features above this threshold are taken to be `significant features of the absorber' and included in the feature mask. The vertical, red lines show the major cuts in the feature mask. For the case of H$_{2}$O, being an absorber/emitter across a broad wavelength range, most wavelengths are included in the feature mask ${\bm \psi}_{H_{2}O}(\lambda)$. CO only absorbs/emits in discrete wavelength ranges and only those will be included in the feature mask of CO, ${\bm \psi}_{CO}(\lambda)$. \label{fig:water+nomask}}
\end{figure}

\begin{figure}[h]
\centering
\includegraphics[width=\columnwidth]{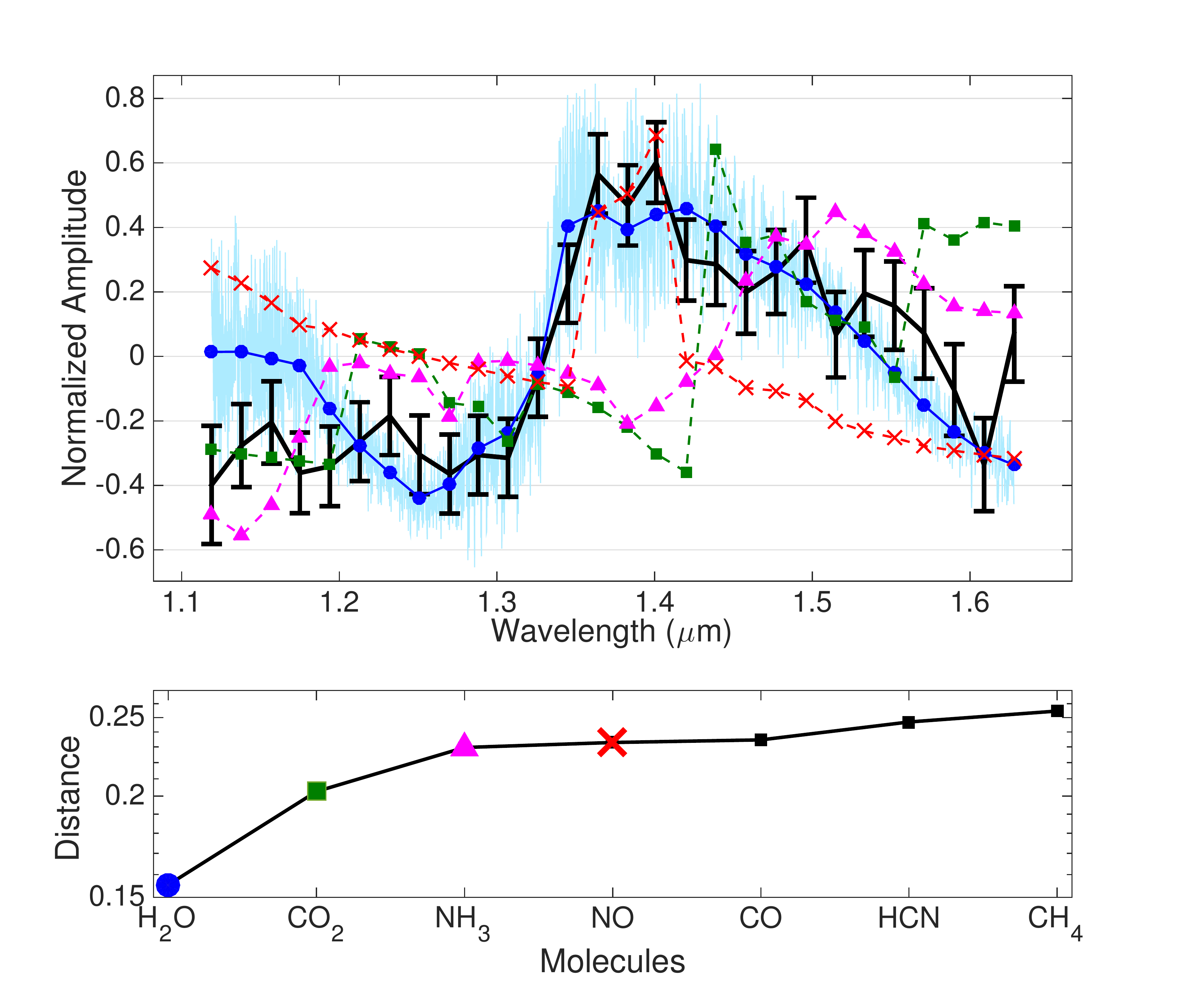}
\caption{TOP: Hubble/WFC3 transmission spectrum of HD 209458b \citep{2013ApJ...774...95D} (black). Overplotted are the principal components of the four best matching molecules binned to the resolution of the data: H$_{2}$O (blue dots), CO$_{2}$ (green squares), NH$_{3}$ (magenta triangles),  NO (red squares). The H$_{2}$O principal component is also shown at a resolution of R~=~1000 (light blue). BOTTOM: The normalised euclidean distance between each individual component and the data. It is clear that H$_{2}$O presents the best match to the data with other molecules being significantly worse.   \label{fig:deming}}
\end{figure}

The purpose of the \marple~module is to constrain the prior space of the Bayesian retrieval in an unbiased way. Given the unknown, varied and complex nature of exoplanet systems it is difficult to pre-suppose atmospheric compositions from `experience'. The most objective approach to atmospheric retrieval of exoplanets would be to assume no prior knowledge at all and to consider all combinations of all atmospheric absorbers known. Whilst desirable, this is computationally infeasible due to the large number of free parameters and often limited spectral resolution of the observed data. The \marple~module attempts to limit the number of possible absorbers by identifying likely molecules in the observed data using a pattern recognition algorithm. It attempts to identify absorption/emission features that are `typical' for a molecular/atomic species and computes the possibility of such indicative features pertaining to a specific molecule compared to all other options.  In this sense, the algorithm is conceptually similar to well established facial recognition algorithms using `eigenfaces' \citep[e.g.][]{Turk:1991eh, 2000SPIE.4067..269C, 2013SPIE.8655E..03G}.

The algorithm is described in the following steps: 

\begin{enumerate}
\item The \marple~module generates a library of atmospheric spectra (equation~\ref{equ:atmabsorb}) for each atmospheric species using the available absorption cross-sections, $\varsigma_{m,\lambda}(T)$. For each species, $m$, spectra are produced  for a large range of atmospheric temperatures, $T$, and mixing ratios $\chi$ (typically $1\times10^{-8} \leq \chi \leq 1\times10^{-1}$ and $500K \leq T \leq 2000K$). Figure~\ref{fig:water+nospectra} shows the transmission spectra of water and CO over a range of temperatures and compositions. Where the molecular absorptions are strongest so are the variations. 
Note that a temperature range can be set for computational efficiency. 

\item  Characteristic spectral features for an atmospheric species vary significantly over the temperature and mixing ratio ranges computed above. These features are key to the identification of the absorber/emitter. 

We can capture these significant variations using a principal component analysis \citep[PCA,][]{Jolliffe:2007vh} where the first component typically indicates the wavelength range over which these features are most prominent (i.e. the amplitude of the variation) and the second component reflects the modulation on the bulk variation, i.e. the features' morphology. 
Hence, for each atmospheric species, we compute the first and second principal components (PCs) over the range of spectra produced above using a single-value-decomposition (SVD).

The SVD of the column vector of spectra ${\bm \alpha}$ is given by 
\begin{equation}
{\bm \alpha}_{m}(T,\chi) = \text{\bf U}{\bm \Sigma} \text{\bf V}^{\text{T}}|_{m}
\label{equ:pre-tau}
\end{equation}

\noindent where $\bf{U}$ and $\bf{V}$ are the left and right unitary matrices respectively and ${\bm \Sigma}$ is the diagonal eingenvalue matrix. Due to the large size of the matrices involved, we approximate equation~\ref{equ:pre-tau} with a randomised, truncated SVD algorithm \citep{Halko:2011bk, Martinsson:2011hb}. The individual principal component is then given by 

%\begin{equation}
\begin{align}
\label{equ:pre-pca}
{\bf pc}_{n,m} &= \text{\bf U}_{n}{\bm \Sigma}_{n} |_{m} \\\nonumber
&= {\bm \alpha}_{m}(T,\chi) \text{\bf V}_{n,m}
\end{align}
%\label{equ:pre-pca}
%\end{equation}

\noindent where $n$ is the PC index. Figure~\ref{fig:waterprincipal} shows the first (blue) and second (red) principal components of water. In this case spectral features are preserved in both components. 

\item We now use the first principal component calculated above to create the `feature mask',  ${\bm \psi}_{m}(\lambda)$, for a given species. This masking guarantees that only wavelengths regions where a given molecule absorbs/emits are correlated against the observed spectrum. The feature mask is given by

\begin{equation}
\label{equ:pre-boolean}
{\bm \psi}_{m}(\lambda) =
\begin{cases}
1 & \text{if } \frac{{\bf pc}_{1,m} - \text{arg min}({\bf pc}_{1,m})}{\text{arg max}({\bf pc}_{1,m})} > \eta \\
0 & \text{otherwise}
\end{cases}
\end{equation}

\noindent where `arg min' and `arg max' stand for the minimal and maximal values of the argument or array. This boolean mask identifies at which wavelengths the characteristic spectral features are stronger than the threshold parameter $\eta$. In other words, the $\eta$ parameter sets the threshold between `molecule present' and `molecule absent' over a given wavelength range. 
We find $\eta = 0.2$ to be a good choice for data with broad wavelength coverage. For spectra consisting of very few ($<$ 20) data points over a narrow wavelength range, the user may not want to exclude (i.e. mask) any data points from the analysis. In such cases setting $\eta$ to a low value, e.g. $\eta = 0.05$, will effectively prevent any wavelength range masking. A range of \marple~module sensitivities can be explored by leaving $\eta$ as free parameter, within user specified limits, over which the \occam~module (section~\ref{sec:occam}) can iterate. 

Figure~\ref{fig:water+nomask} shows the creation of the `feature mask' for water and CO. Wavelengths with the normalised ${\bf pc}_{1,m}$ bigger than $\eta$ will be included into the feature mask of molecule $m$. 

\item For each species we now convolve the second PC and the observed data ${\bf x}$ with the `feature mask' to obtain the masked PC and observed data vectors $\widehat{\bf pc}_{2,m}$ and $\hat{\bf x}_{m}$ respectively

\begin{equation}
\label{equ:pre-pcamasked}
\widehat{\bf pc}_{2,m} = {\bm \psi}_{m} \otimes {\bf pc}_{2,m}
\end{equation}

\begin{equation}
\label{equ:pre-masked}
\hat{\bf x}_{m} = {\bm \psi}_{m} \otimes \bar{\bf x}
\end{equation}

\noindent where $\otimes$ denotes the convolution operator and $\bar{\bf x}$ is the normalised observed data vector given by

\begin{equation}
\bar{\bf x} = \left ( \frac{\bf x - \text{arg min}(\bf x)}{\text{arg max}(\bf x)} \right )
\label{equ:norm-data}
\end{equation}

\item In order to select the set of best matching molecular/atomic species to the observed spectrum, we implement a variant of the K-nearest-neighbour \citep[K-NN,][]{Cover:1967jq, Altman:1992kb} algorithm based on the euclidian distance between the spectral library principal component vectors and the observed data.  For this we calculate the $L^{2}$ norm (also known as Euclidean norm or Euclidean distance), $\mathfrak{d}_{m}$,  between masked data and the masked second PC for each molecule, $m$ 

\begin{equation}
\label{equ:pre-l2}
\mathfrak{d}_{m} = \frac{1}{N} ||\hat{\bf x}_{m} - \widehat{\bf pc}_{2,m}||_{2}
\end{equation}

\noindent where $N$ is the total number of data points in ${\bf x}$.  We now sort the euclidian distances in ascending order to form the monotonically increasingf sequence of $\mathfrak{d}_{m}$

\begin{equation}
\label{equ:pre-fd}
f(\mathfrak{d}_{m}) = \{\mathfrak{d}_{1,m}, \mathfrak{d}_{2,m}, \dots, \mathfrak{d}_{\phi,m}\}%~~~\text{where}~|\mathfrak{d}_{\varphi-1}| < |\mathfrak{d}_{\varphi}|
\end{equation}

\noindent where $\phi$ is the sequence index and $|\mathfrak{d}_{\phi-1}| < |\mathfrak{d}_{\phi}|$.  The total distance is given by $\mathfrak{d}_{total} = \sum_{\phi}^{M} \{\mathfrak{d}_{\phi} \}$, where M is the total number of molecules considered. The algorithm distinguishes between the cluster of best matching  (low $\mathfrak{d}$) and worst matching species (high $\mathfrak{d}$) by finding the series index associated with the highest second derivative of $f(\mathfrak{d}_{m})$

\begin{equation}
\label{equ:pre-cluster}
\varphi = \phi ~ \text{where}~ \left [\text{arg max} \left ( \frac{\text{d}^{2}f}{\text{d}\mathfrak{d}^{2}} \right )\right ]
\end{equation}

\noindent the number of molecules selected by the preprocessor, $N_{m}$, is then given by 

\begin{equation}
N_{m} =
\begin{cases}
\varphi &\text{if } \varphi > N_{m, min}\\
N_{m,min} &\text{otherwise}
\end{cases}
\end{equation}

\noindent where $N_{m,min}$ is a minimal number of molecules to be selected which can be set by the user. The selected molecules are given by $m_{select} = m_{\phi < \varphi}$.

\end{enumerate}

Once determined the \marple~module passes its list of selected atmospheric species to the central data module (see figure~\ref{fig:flowchart}) which will prepare all inputs for further analysis. The efficiency of the \marple~module is a function of spectral resolution and signal to noise (S/N) of the data. This is self evident as any identification of features is impaired by a too coarse wavelength grid (R $<$ 10) or high noise levels (S/N $<$ 5). For those extreme cases, the user may specify a list of `must include' molecules in the parameter files to be considered by the \taurex.  

\subsection{Example: HD 209458b}

We demonstrate the \marple~module using a transmission spectrum of the hot-Jupiter HD~209458b obtained by the {\it Hubble}/WFC3 camera \citep{2013ApJ...774...95D}. The top of figure~\ref{fig:deming} shows the transmission spectrum in black and the principal components of the four best matching molecules. Note that all amplitudes are normalised and we only compare morphologies. The bottom panel summarises the normalised Euclidean distances (equation~\ref{equ:pre-l2}) for individual molecules. Here the black continuous line represents the sequence $f(\mathfrak{d}_{m})$ in equation~\ref{equ:pre-fd}. The \marple~module returns water as the most likely molecule present with CO$_{2}$ a more distant second. The presence of water as main absorbing species is in good agreement with the results of previous analyses \citep{2013ApJ...774...95D, 2014ApJ...791L...9M}.

\section{Retrieval}
\label{sec:retrieval}

%As described in sections~\ref{sec:linelists} \& \ref{sec:marple}, the absorption cross-section line-lists for $m_{select}$ are read and interpolated onto a fine temperature grid. At each iteration of the retrieval described below, the forward model (section~\ref{sec:forward} selects the appropriate cross-section list for the current temperature considered. Whereas such a nuanced treatment may be excessive for very hot extrasolar planets, the importance of temperature dependent cross-sections can only be stressed for warm ($\sim 400 - 1000$K) planets. 

\taurex~features three independent retrieval methods: 1) least-square minimisation using a quasi-Newtonian Limited-Memory Broyden-Fletcher-Goldfarb-Shannon (LM-BFGS) algorithm, section~\ref{sec:leastsquare}, 2) an Adaptive, multi-chain Markov Chain Monte Carlo algorithm, section~\ref{sec:mcmc} and 3) a nested-sampling algorithm using MultiNest, section~\ref{sec:nested}. 

Programatically, individual minimisation routines submit standardised requests to the central data module which in turn handles all calls to the forward module, the T-P profile and required inputs (figure~\ref{fig:flowchart}). This modular approach guarantees consistency between model, data and retrieval codes as well as a high degree of flexibility in the analysis of the observed data.

%\section{Minimization \& Sampling}
%\label{sec:min}

\subsection{Prior bounds}

\taurex~by default uses uniform priors for all free parameters. As default, the isothermal temperature bounds are T$_{equ} \pm 200$ K, where T$_{equ}$ is the planetary equilibrium temperature (this can either be derived by \taurex~given planetary/orbital parameters or set by the user). The molecular mixing ratios are bounded between 0.0 - 1.0$\times 10^{-1}$ by default. All prior bounds can be manually specified by the user. The planet-star ratio, $(R_{p} /R_{\ast})^{2}$, is treated as free parameter by default, with its upper/lower bounds derived from the reported observational uncertainty on this ratio. \citet{2014RSPTA.37230086G} and \citet{Benneke:2013vx}, amongst others, have noted strong degeneracies between the value of $(R_{p} /R_{\ast})^{2}$ and various retrieval parameters (e.g. H$_{2}$O abundance and cloud opacities). We will further explore these degeneracies in a subsequent publication (Rocchetto et al. in prep.).

\subsection{LM-BFGS minimization}
\label{sec:leastsquare}

The least-square minimisation allows us to obtain a quick look at the optimal model fit for the data without using the computationally expensive MCMC or Nested sampling routines. In this respect it is key to the pre-burning of the MCMC chain, as at least one chain can be started at the optimal solution and hence does not require a burn-in time \citep{Brooks:2011via}, as well as providing a valuable consistency check between model fits produced by the MCMC and MultiNest routines.

Large numbers of free parameters are often a limiting factor for simplex-downhill algorithms \citep[e.g.][]{Nelder:1965in} commonly used. We find such amoeba algorithms to be insufficient and to often get stuck in local minima. \taurex~uses the LM-BFGS \citep{Zhu:1997wv,Morales:2011em} algorithm instead, which being quasi-Newtonian uses the inverse Hessian matrix of the $\chi^{2}$ surface to efficiently and robustly converge to the global maximum. We furthermore find the LM-BFGS to be more robust in the presence of observational noise than comparable methods. 

\newpage
\subsection{Bayesian analysis}
\label{sec:Bayesian}

The Bayesian argument is given by 

\begin{equation}
\label{equ:bayesequation}
P(\theta | {\bf x}, \mathcal{M}) = \frac{P({\bf x} | \theta, \mathcal{M}) P(\theta, \mathcal{M})}{P({\bf x} | \mathcal{M})}
\end{equation}

\noindent where $P(\theta, \mathcal{M})$ is the Bayesian prior which we take to be uniform throughout this paper. The number and type of absorbing species, as well as the equilibrium temperature of the planet defining the forward model, $\mathcal{M}$, are set by the \marple~module (section~\ref{sec:marple}). $P(\theta | {\bf x}, \mathcal{M})$ is the posterior probability of the model parameters $\theta$ given the data, ${\bf x}$ assuming the forward model $\mathcal{M}$. The likelihood, $P({\bf x} | \theta, \mathcal{M})$ is given by the Gaussian 

\begin{equation}
P({\bf x} | \theta, \mathcal{M}) = \frac{1}{\varepsilon \sqrt{2\pi}} ~\text{exp}\left [{-\frac{1}{2}\sum_{\lambda}^{N}\left ( \frac{x_{\lambda} - \mathcal{M}_{\lambda} }{\varepsilon_{\lambda}} \right )^{2}} \right ]
\label{equ:normal-likelihood}
\end{equation}

\noindent where $\epsilon$ is the error on the observed spectral point. As opposed to the nested sampling described in the next section, an MCMC does not sample the Bayesian partition function (also known as Bayesian Evidence) and equation~\ref{equ:bayesequation} reduces to 

\begin{equation}
\label{equ:mcmc}
P(\theta | {\bf x}, \mathcal{M}) \propto P({\bf x} | \theta, \mathcal{M}) P(\theta, \mathcal{M}).
\end{equation}

\subsubsection{MCMC}
\label{sec:mcmc}

%\begin{equation}
%\label{equ:bayesequation}
%P(\theta | {\bf x}, \mathcal{M}) \propto P({\bf x} | \theta, \mathcal{M}) P(\theta, \mathcal{M})
%\end{equation}

MCMC routines are commonly used in the field of extrasolar planets  \citep[e.g.][]{ford06,burke10,bakos07,knutson07a,cameron07,charbonneau09, bean10, Kipping:2011fl,gregory11,2012ApJ...761....7C,2014Natur.505...69K,Braak:2006js,2013ApJ...775..137L,2014ApJ...791L...9M,2012ApJ...753..100B,TerBraak:2008iw,ForemanMackey:2013cf,Goodman:2010et, 2014ApJ...791L...9M}. \taurex~provides an implementation of the Delayed-Rejection Adaptive-MCMC \citep[DRAM,][]{Haario:2006hy}. We refer the interested reader to the cited literature and here only provide a brief overview. The DRAM algorithm differs from a more classical Metropolis-Hastings sampler \citep{Metropolis:1953vj, Hastings:1970wm, Brooks:2011via} in two aspects: 1) It implements a delayed rejection algorithm and 2) an adaptive proposal distribution calibrated using the covariance of the sample path of the MCMC chain. For additional information on DRAM, we refer the reader to Appendix~\ref{appendix:dram} and the relevant literature.  

\taurex~runs several MCMC chains in parallel to check convergence and increase the sampling of the likelihood space. The number of chains is user defined but usually set to 4-5 and limited by the number of available CPUs. The first primary chain is started at the optimal values determined by the LM-BFGS, avoiding significant burn-in time \citep{Brooks:2011via}. All secondary chains' starting positions are offset from the optimum by a random distance and direction of at least 10$\%$ of the prior width. These secondary chains are run with a burn-in period of typically 10$\%$ of the total chain length. Burn-in and chain lengths are user defined.   

\subsubsection{Nested Sampling}
\label{sec:nested}

Nested sampling (NS) algorithms \citep{2004AIPC..735..395S, Skilling:2006wf, 2006ApJ...638L..51M, Chopin:2010vy, 2011MNRAS.414.1418K,  2005AIPC..803..189J} are becoming increasing popular in extrasolar planets \citep[e.g.][]{Kipping:2012uz, Placek:2013wu} as well as \citet{Benneke:2013vx} for atmospheric retrieval. Here we include an implementation of \multinest~\citep{2008MNRAS.384..449F, Feroz:2009jn,Feroz:2013wp}. MCMC algorithms are commonly used for parameter estimation by solving equation~\ref{equ:mcmc}. Whereas MCMC explores the likelihood space by means of a Markovian chain, NS performs a general Monte Carlo (MC) analysis which is periodically constrained by ellipsoids encompassing spaces of highest likelihoods. Note that unlike MCMC, NS does not depend on a pre-determined proposal density and can hence better explore highly degenerate and non-Gaussian regimes. 
Using NS, we can compute the Bayesian evidence (or simply evidence), which is given by the integral required to normalising equation~\ref{equ:mcmc}

\begin{equation}
\label{equ:bayesevidence}
E = \int P({\bm \theta} | \mathcal{M}) P({\bf x} | {\bm \theta}, \mathcal{M}) \text{d}{\bm \theta}
\end{equation}

\noindent where $E = P({\bf x} | \mathcal{M})$ is the evidence. The evidence allows us to test the adequacy of the model itself and to perform model selection as described in the following section. Posterior distributions for parameter estimations are returned as by-product of \multinest~ and should be similar to posteriors obtained by the MCMC. Note that through the very different sampling techniques and fewer constraints on the proposal density for NS, we expect MCMC posteriors to be a `smoothed' version of the NS's. \taurex~allows the choice between importance nested sampling (INS) and the more classical NS. Through the sampling process, INS retrains all accepted as well as rejected proposal points which allows for a more accurate integration of the evidence \citep{Feroz:2013wp}. Nested Sampling (in its \multinest~implementation) is highly efficient and easily parallelisable, allowing an easy scaling to cluster computing. We here use the NS approach as our main means of retrieval with the MCMC implementation providing a valuable cross check on the final results.

\section{Model selection}
\label{sec:occam}

For an inverse retrieval problem, such as the one discussed here, the idea of model selection is highly relevant but rarely discussed due to the computational expense and complexities involved. Notable examples of Bayesian model selection in atmospheric retrieval are \citet{Benneke:2013vx,2013ApJ...778..183L, 2014ApJ...784..133S}. Here we explicitly make the distinction between optimal estimation of parameters and the adequacy of the parameter and/or model itself. 

We perform two tests after each \taurex~run:

\begin{enumerate}
\item \underline{Parameter adequacy:} is a parameter (e.g. a given molecular species) required to describe the underlying physics? If not, is the model considered {\it over-complete}? In the case of {\it over-completeness} the forward model may be too complex (not obeying Occam's razor). This can lead to overfitting in the worst case or in the best case a reduction in retrieval efficiency. 
\item \underline{Model adequacy:} are parameters missing in the model considered, i.e. is the model {\it under-complete}? In the case of model {\it under-completeness} the data is better accounted for by a more complex model. For example, a cloudy exoplanetary atmosphere cannot be modelled adequately by a cloud-free atmospheric model. Here the presence of clouds could force a cloud-free model to compensate for the extra absorption using molecular/atomic absorbers. This introduced bias often cannot be discerned from parameter estimating algorithms such as MCMC, maximum likelihood and similar methods.
\end{enumerate}

Determining a model that is adequate to the data's complexity is hence paramount. 

\taurex~tries to perform model selection in an intelligent way through the \occam~module. The \occam~ module will perform model selection until a complete model is determined. It will iterate through models, appropriately increasing or decreasing the model complexity through interaction with the \marple~module (see figure~\ref{fig:flowchart}, point 4).

\subsection{Over-complete models}

The over-complete model features an unnecessary complexity. Here the desired model is a sub-set of the more complex model initially run. We referred to these models as being `nested'. Complexity in parametric models (such as the forward models of atmospheric retrieval) is usually synonymous with number of free-parameters. Hence we can define the nested model $ \mathcal{M}_{{\bm \theta}-\theta_{\gamma}}$ as sub-set of the more complex one $ \mathcal{M}_{\bm \theta} $,

\begin{equation}
\left. \mathcal{M}_{{\bm \theta}-\theta_{\gamma}} = \mathcal{M}_{\bm \theta} \right | _{\theta_{\gamma}= 0}
\label{equ:nested}
\end{equation}

\noindent where ${\bm \theta}$ is a column-vector of all model parameters and $\theta_{\gamma}$ is an individual parameter. The Bayes factor (see section~\ref{sec:undercomplete}) allows us to perform this model selection by marginalising out individual parameters \citep{Benneke:2013vx,2014ApJ...784..133S}. For `nested'  models we can derive the simpler to compute Savage-Dickey density ratio (SDR) \citep{Dickey:1971vm, Verdinelli:1995vb,Marin:2010um, 2013JCAP...09..013V} 

\begin{equation}
SDR = \left. \frac{P(\theta_{\gamma} |  {\bf x}, \mathcal{M}_{{\bm \theta}})}{P(\theta_{\gamma} | \mathcal{M}_{\bm \theta}) }\right|_{\theta_{\gamma} = 0} 
\label{equ:savage}
\end{equation}

%\begin{equation}
%\frac{E_{1}}{E_{2}} = \left. \frac{P(\theta_{\gamma} |  {\bf x}, \mathcal{M}_{{\bm \theta}})}{P(\theta_{\gamma} | \mathcal{M}_{\bm \theta}) }\right|_{\theta_{\gamma} = 0} 
%\label{equ:savage}
%\end{equation}

\noindent where $P(\theta_{\gamma} |  {\bf x}, M_{{\bm \theta}})$ is the marginalised posterior of $\theta_{\gamma}$ and $P(\theta_{\gamma} | M_{\bm \theta})$ its respective prior distribution. A comprehensive derivation and discussion of equation~\ref{equ:savage} can be found in \citet{2013JCAP...09..013V}. This ratio of densities at $\theta_{\gamma} = 0$ is indicative of whether a simpler model not containing $\theta_{\gamma}$ is sufficient or whether a more complex model is preferred by the data. To assess the significance of the evidence towards a complex rather than a simpler model, we compare the outcome of equation~\ref{equ:savage} to the Jeffrey's scale \citep{Jeffreys:1998tt}. We adopt a slightly modified version of \citet{Kass:1995vb} in table~\ref{tbl:sdr}

\begin{table}[h]
%\center
\caption{ Jeffrey's scale for model selection} \label{tbl:sdr}
\begin{tabular}{ll}
\hline\hline
2ln($SDR$) & Preference for simplified model $\mathcal{M}_{{\bm \theta}-\theta_{\gamma}}$ \\
  \hline
$>$ 10 & Very strong preference for excluding $\theta_{\gamma}$ \\
10  to 6 & Strong preference for excluding $\theta_{\gamma}$  \\
6 to 2 & Substancial preference for excluding $\theta_{\gamma}$ \\
2 to 0 & Insignificant preference for excluding $\theta_{\gamma}$ \\
\hline 
\hline
 & Preference for complex model $\mathcal{M}_{{\bm \theta}}$ \\
  \hline
 0 to -2 & Insignificant preference for including $\theta_{\gamma}$ \\
-2 to -6 & Substancial preference for including $\theta_{\gamma}$ \\
-6  to -10 & Strong preference for including $\theta_{\gamma}$  \\
$<$ -10 & Very strong preference for including $\theta_{\gamma}$ \\

 \end{tabular}
\end{table}

The \occam~module calculates the Savage-Dickey ratio for each model parameter and adjust the model complexity accordingly in case of a strong preference for a simpler forward model. 

\subsection{Under-complete models}
\label{sec:undercomplete}

Should the model at hand not be over-complete, the \occam~ module tests for model under-completeness, i.e. is the model complex enough. For this we iteratively re-run the retrieval process allowing the \marple~module to add the two next most likely molecular opacities to the current selection of opacities. We then compute the global model evidence, $E$, and compute the  Bayes factor \citep{Kass:1995vb,Weinberg:2012vj}. The Bayes factor is given by the ratio of model probabilities $P(\mathcal{M}_{1} | {\bm \theta})$

\begin{equation}
\label{equ:bayesfactor1}
\frac{P(\mathcal{M}_{2} | {\bm \theta})}{P(\mathcal{M}_{1} | {\bm \theta})} = \frac{P(\mathcal{M}_{2})}{P(\mathcal{M}_{1})}\frac{P({\bf x}|\mathcal{M}_{2})}{P({\bf x}|\mathcal{M}_{1})} = \frac{P(\mathcal{M}_{2})}{P(\mathcal{M}_{1})}\frac{E_{2}}{E_{1}}
\end{equation}

\noindent which can be expressed as fraction of the Evidences and the prior distribution of the models. Most times we can assume the model priors to be identical $P(\mathcal{M}_{1}) =    P(\mathcal{M}_{2})$, reducing equation~\ref{equ:bayesfactor1} to 

\begin{equation}
\label{equ:bayesfactor}
E_{2}/E_{1} = \frac{P({\bf x} | \mathcal{M}_{2})}{P({\bf x} | \mathcal{M}_{1})} .
\end{equation}

\noindent Using the Jeffrey's scale the \occam~module determines whether an improvement to the fit is achieved using a more complex model.

\section{Outputs}
\label{sec:output}

The output module generates the best fit transmission model, plots of all marginalised and conditional posteriors as well as statistics on individual parameters and model adequacy. Examples of these outputs can be found in the following section.

\section{Example}
\label{sec:example}

\begin{figure}[h]
\centering
\includegraphics[width=\columnwidth]{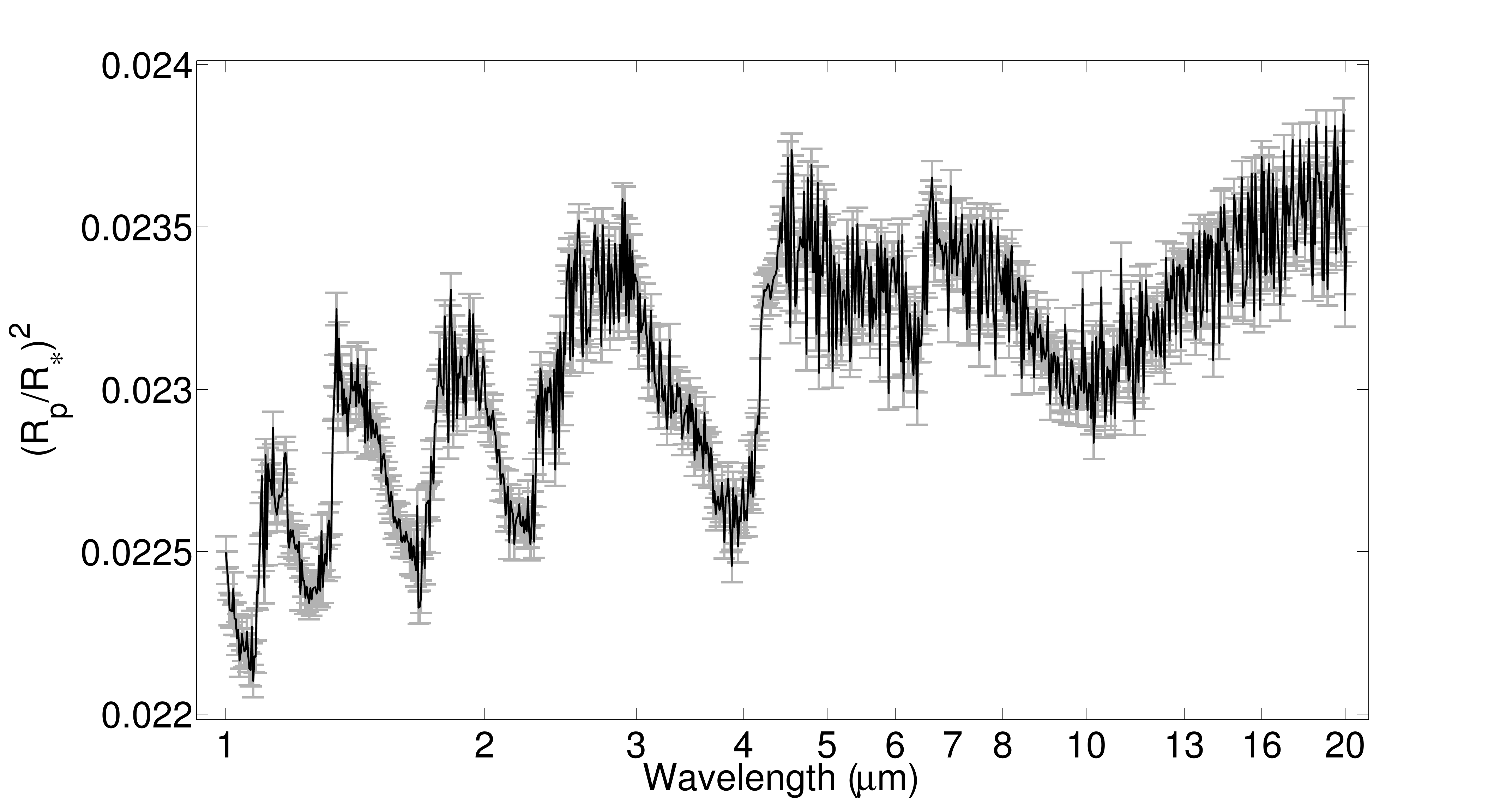}
\caption{Simulated example spectrum of a carbon-rich hot-Jupiter used in section~\ref{sec:example}. Temperature and abundances of the main absorbers, H$_{2}$O, CO, CO$_{2}$, NH$_{3}$ and CH$_{4}$ are given in table~\ref{inputtable}. The bulk planet/star and orbital properties are based on the hot-Jupiter HD209458b.  \label{fig:spectrum}}
\end{figure}

In this section we demonstrate the output of \taurex~using a simulated hot-Jupiter. We base the simulation on a HD209458b like planet/star system \citep{charbonneau00,2010MNRAS.408.1689S} with temperature and bulk composition taken from \citet{2014A&A...562A..51V}. We choose a wavelength range of $1 - 20 \mu$m at a constant resolution of R = 300 and constant error bars of 50ppm. Table~\ref{inputtable} summarises the inputs and figure~\ref{fig:spectrum} shows the input spectrum to \taurex. Whilst such an example may be optimistic given currently available data we would like to note the following: 1) In order to demonstrate the retrieval accuracy of \taurex~one needs a precise data set, 2) Future observatories and missions (e.g. JWST, E-ELT and dedicated missions) will yield data of comparable or better quality over broad wavelength ranges. 

\begin{table*}
\centering
\caption{Model input and retrieval results for LM-BFGS, MCMC and NS. All values (but temperature) are in units of fractional column density.} \label{inputtable}
\vspace{10pt}
\begin{tabular}{r  l  l  l  l l}
\hline\hline
Parameters & Model & LM-BFGS  & MCMC & Nested Sampling \\ \hline
Temp. (K) & 1400 & 1419.14             &  1403.30 $\pm$ 9.88                                         & 1403.87 $\pm$ 9.26\\
H$_{2}$O & 2$\times$10$^{-3}$  &1.94$\times$10$^{-3}$&  1.90$\times$10$^{-3 }$ $\pm$   3.27$\times$10$^{-5}$    & 1.90$\times$10$^{-3 }$ $\pm$   3.11$\times$10$^{-5}$ \\
CH$_{4}$ & 2$\times$10$^{-6}$  &2.04$\times$10$^{-6}$&  2.25$\times$10$^{-6}$ $\pm$   1.45$\times$10$^{-6}$ & 2.17$\times$10$^{-6 }$ $\pm$   1.42$\times$10$^{-6}$ \\
CO             & 2$\times$10$^{-3}$  &1.95$\times$10$^{-3}$&  1.97$\times$10$^{-3}$ $\pm$   1.26$\times$10$^{-4}$   & 1.97$\times$10$^{-3}$ $\pm$   1.22$\times$10$^{-4}$\\
CO$_{2}$ & 2$\times$10$^{-5}$  &2.41$\times$10$^{-5}$& 2.48$\times$10$^{-5 }$ $\pm$   2.59$\times$10$^{-6}$    & 2.48$\times$10$^{-5 }$ $\pm$   2.60$\times$10$^{-6}$ \\
NH$_{3}$ & 2$\times$10$^{-7}$  &1.48$\times$10$^{-6}$& 1.18 $\times$10$^{-6 }$ $\pm$   9.69$\times$10$^{-7}$    & 1.18$\times$10$^{-6 }$ $\pm$   9.69$\times$10$^{-7}$ \\
\hline
 \end{tabular}
\end{table*}

The data is passed through \taurex~as described in the previous sections. The \marple~module suggested the correct molecules as potentially important absorbers given the data and their wavelength ranges. In addition to the molecules listed in table~\ref{inputtable}, it also identified H$_{2}$C$_{2}$ as possible absorber which was subsequently rejected by the \occam~ module and the transmission module was updated to reflect the true model of the data.  

\begin{figure}
\centering
\includegraphics[width=\columnwidth]{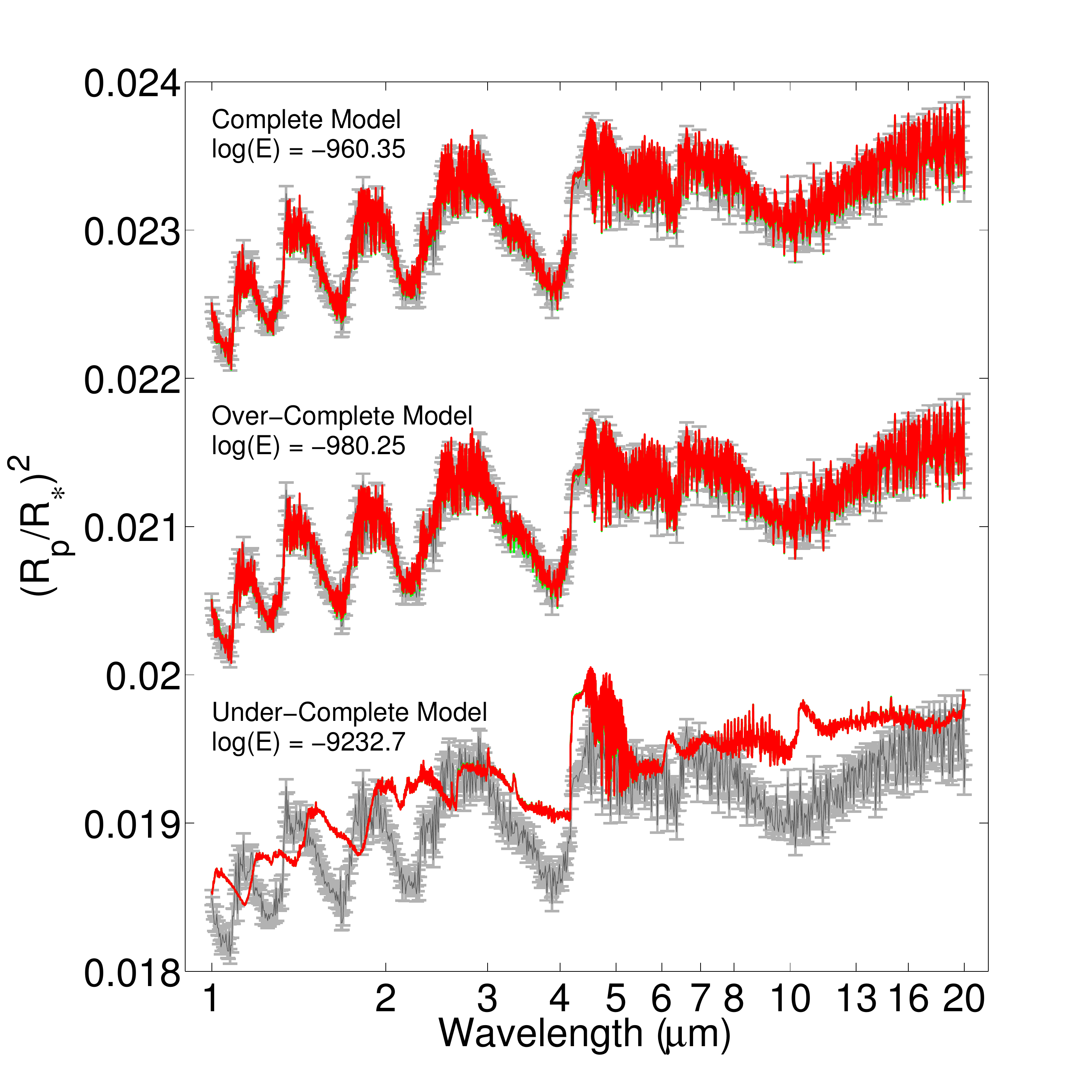}
\caption{showing the best fitting models, in red, for the complete (top), over-complete (middle) and under-complete (bottom) model cases as described in section~\ref{sec:example}. The best fitting models are offset along the ordinate for clarity and over-plotted on the `observed' spectrum in grey. The global Bayesian Evidences, log(E), are given for each case, quantifying the adequacy of each model given the data. As expected the evidence strongly favours the correct, complete model.   \label{fig:modelselection}}
\end{figure}

\begin{figure*}
\centering
\includegraphics[width=\textwidth]{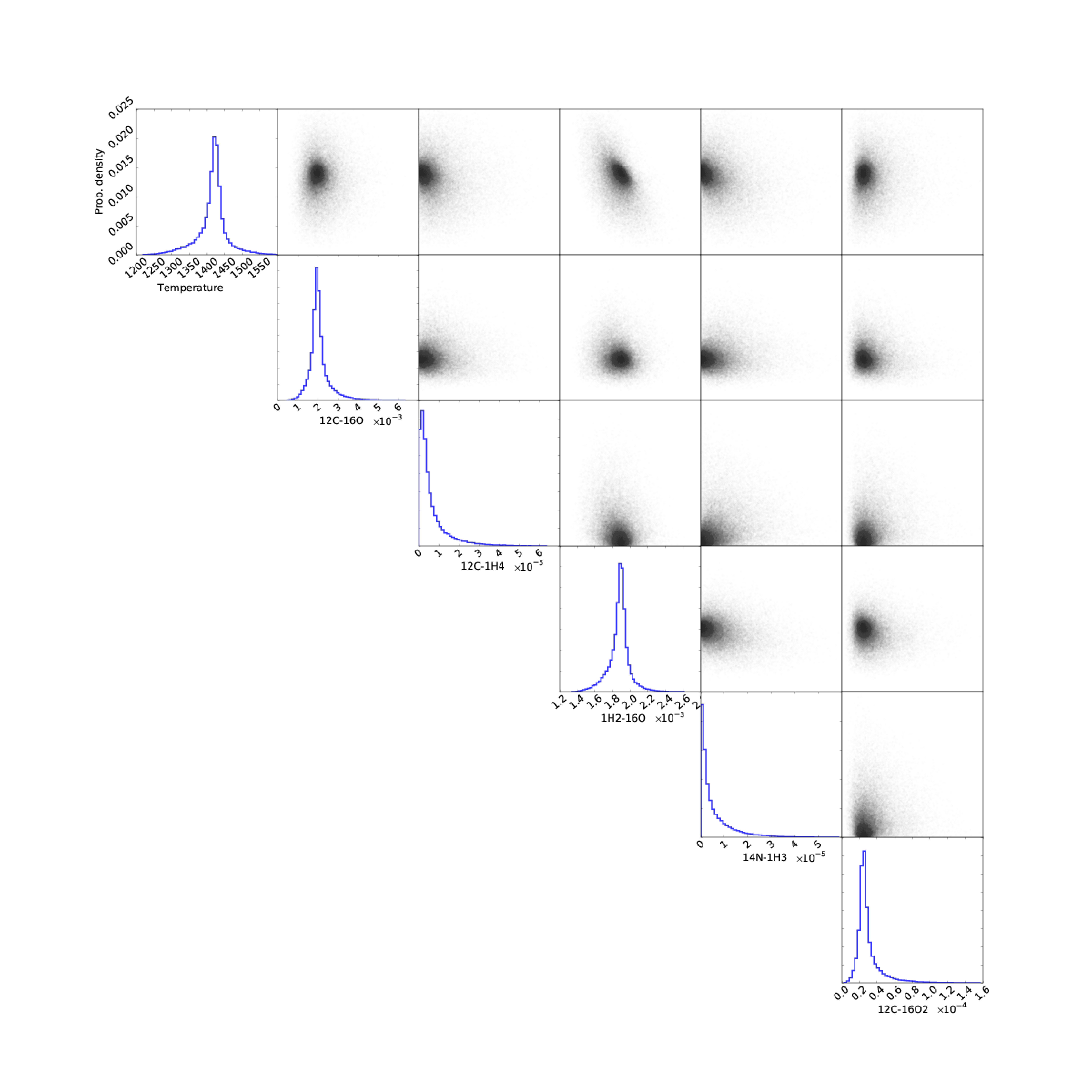}
\caption{showing the marginalised and conditional posterior distributions for the Nested Sampling for the complete model in section~\ref{sec:example}. We find the highest correlation between atmospheric temperature and water absorption. With H$_{2}$O being the strongest absorber across the broadest wavelength range, this correlation between abundance and thermal broadening due to temperature changes is to be expected.  \label{fig:post-nest-complete}}
\end{figure*}

\begin{figure*}
\centering
\includegraphics[width=\textwidth]{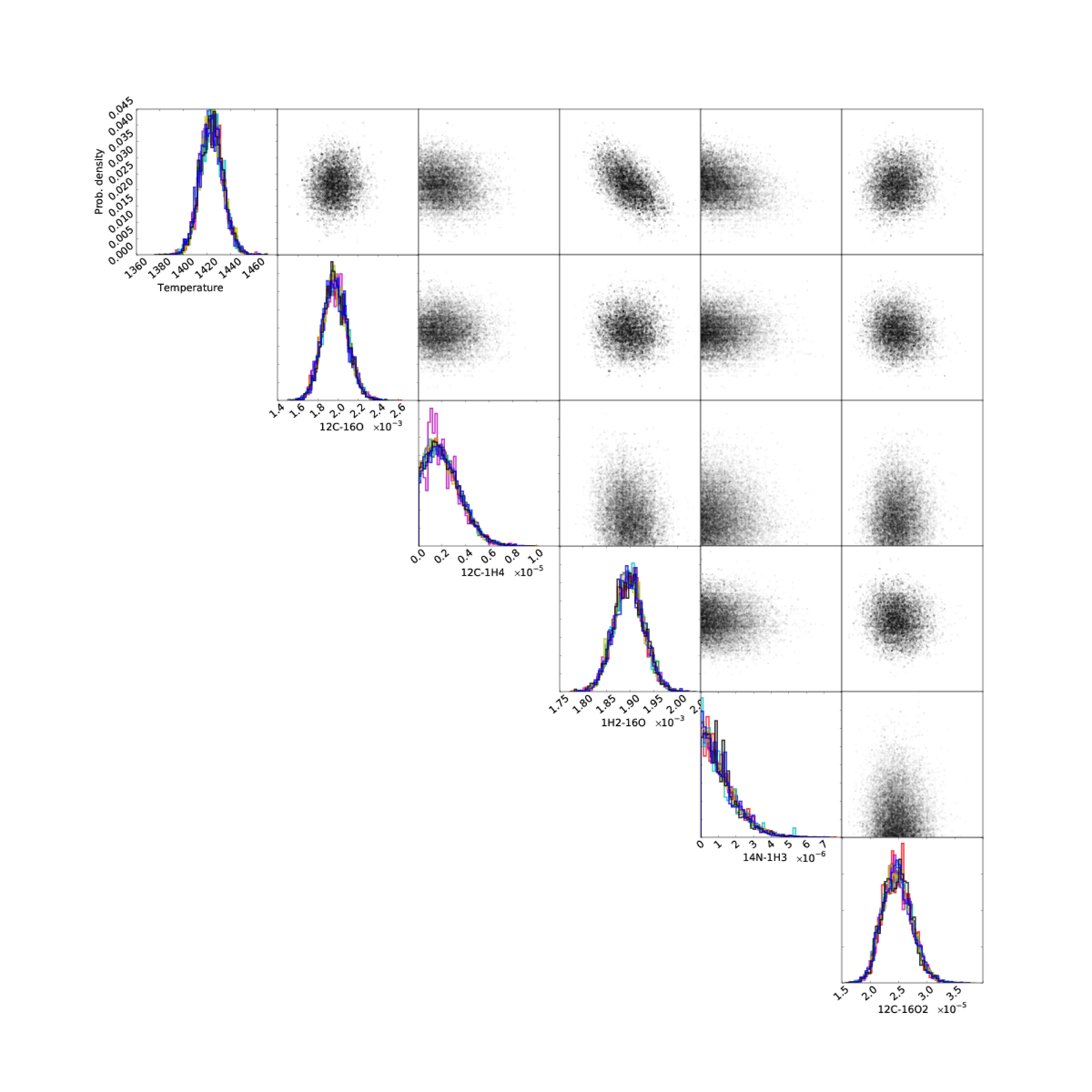}
\caption{showing the marginalised and conditional posterior distributions for the MCMC run of the complete model. Different colours in the marginalised posterior plots represent individual MCMC chains. The very good overlap of these independent chains indicates a good convergence of the code.  \label{fig:post-mcmc-complete}}
\end{figure*}

The retrieved temperature and abundance values for the LM-BFGS, MCMC and Nested Sampling algorithms are summarised in table~\ref{inputtable}. \taurex~does not compute a formal error for the LM-BFGS result as only the MCMC and Nested Sampling results are considered to be final data products. Figure~\ref{fig:modelselection} (top spectrum) shows the best-fit transmission model for the MCMC (green) and the Nested Sampling (red) algorithms. Figures~\ref{fig:post-nest-complete}~\&~\ref{fig:post-mcmc-complete} show the marginalised and conditional posterior distributions for the MCMC and Nested Sampling results respectively. The MCMC results consist of 8 independent chains (note the different colours in the marginalised posteriors representing the results of individual chains) and 2.5$\times$10$^{4}$ samples each, including a 10$\%$ burn-in period. The Nested sampling results used 4000 initial live-points and 8.3$\times$10$^{4}$ replacements.  
Note the Nested Sampling posteriors to be more kurtotic than the MCMC results. This is due to a finer sampling of the maximum likelihood space by the NS.

Table~\ref{inputtable} summarises the retrieved abundances. All major species in the simulated transmission spectrum as well as the isothermal temperature were retrieved with great fidelity by all retrieval methods. Mixing ratios for NH$_{3}$ and CH$_{4}$ were set purposefully low to test the retrievability of molecular abundances at the limits of data uncertainties. As shown in section~\ref{sec:example-model} and table~\ref{table:sdrexample}, these detections were identified as `insignificant' by the \occam~module.   

\begin{figure*}
\centering
\includegraphics[width=\textwidth]{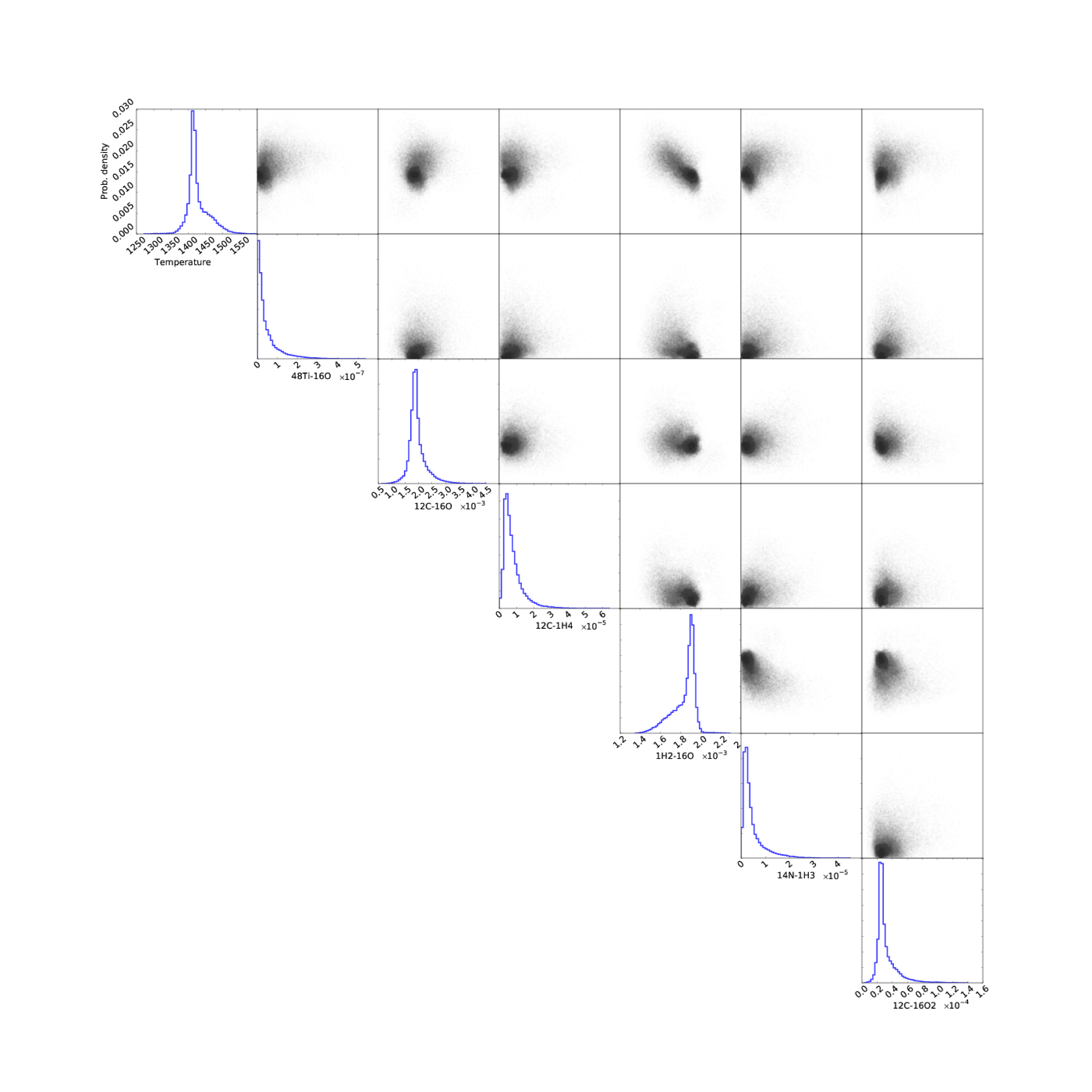}
\caption{showing marginalised and conditionals posterior distributions of the Nested Sampling run for the over-complete model in section~\ref{sec:example-model}. Otherwise identical to figure~\ref{fig:post-nest-complete}. \label{fig:post-nested-overcomplete}}
\end{figure*}

\begin{figure*}
\centering
\includegraphics[width=\textwidth]{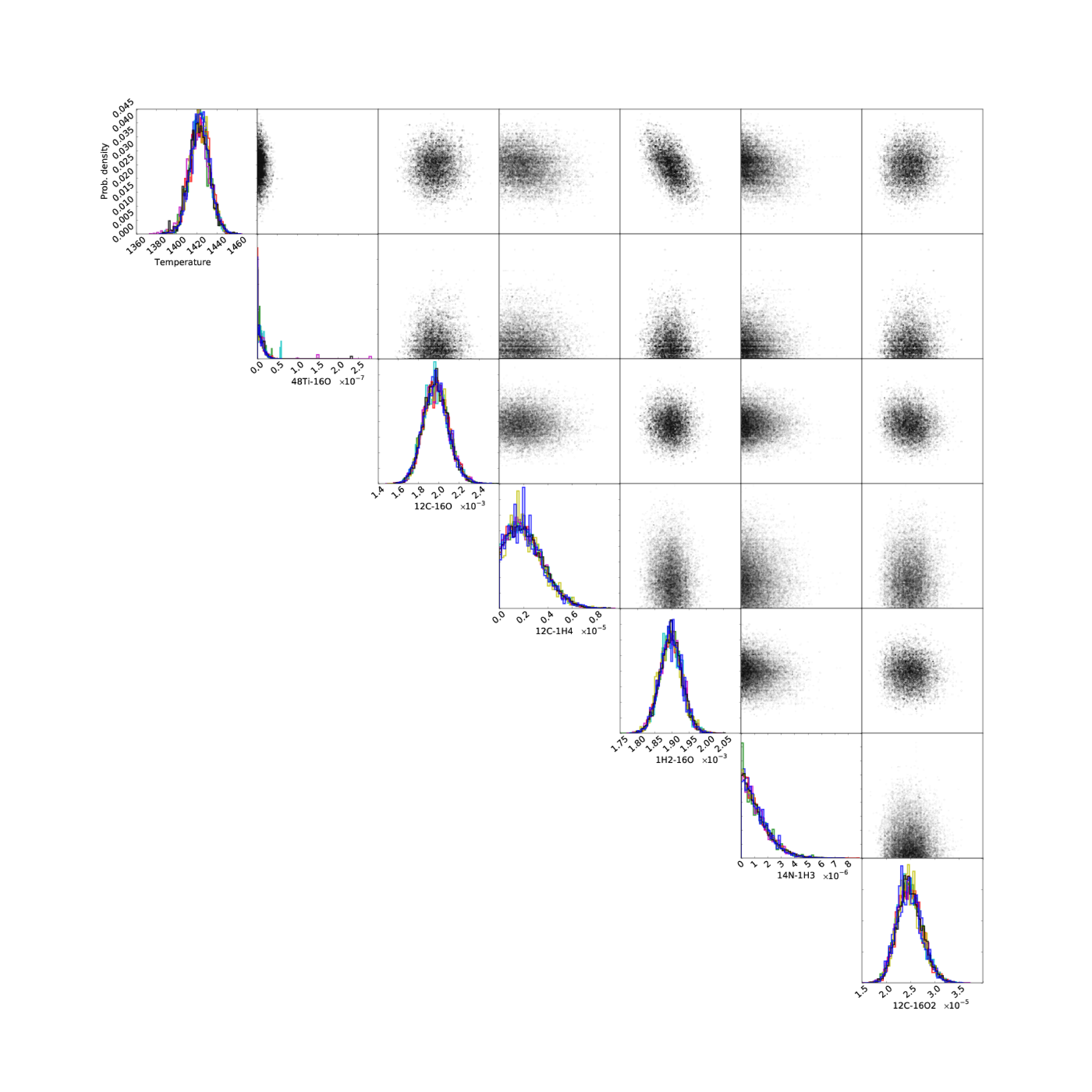}
\caption{showing marginalised and conditionals posterior distributions of the MCMC run for the over-complete model in section~\ref{sec:example-model}. Otherwise identical to figure~\ref{fig:post-mcmc-complete}. \label{fig:post-mcmc-overcomplete}}
\end{figure*}

\begin{figure*}
\centering
\includegraphics[width=\textwidth]{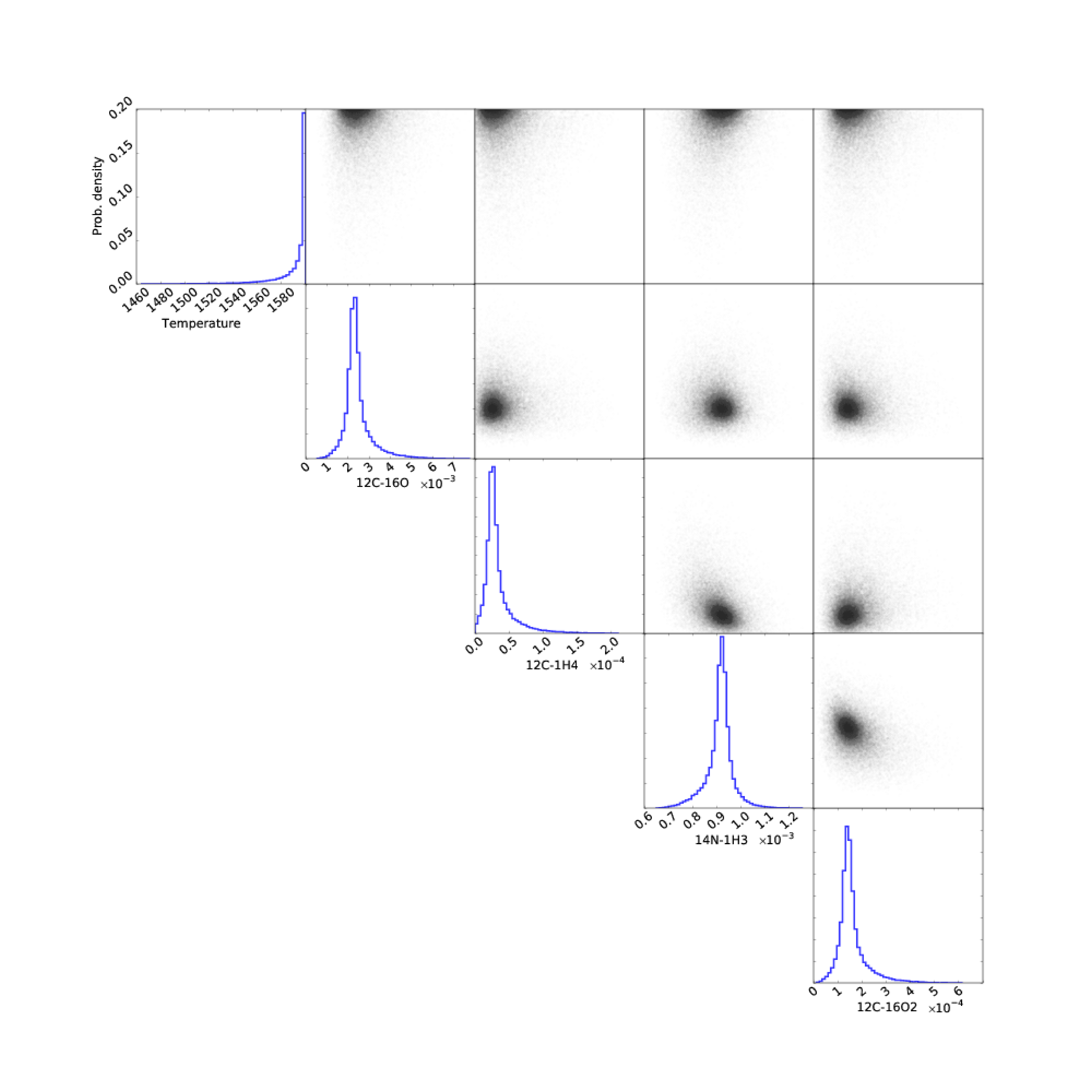}
\caption{showing marginalised and conditionals posterior distributions of the Nested Sampling run for the under-complete model in section~\ref{sec:example-model}. In the absence of the main absorbing species, H$_{2}$O, \taurex~tries to compensate for lacking opacity by increasing thermal broadening. This results in the planetary temperature converging to the upper end of the prior. Otherwise identical to figure~\ref{fig:post-nest-complete}. \label{fig:post-nested-undercomplete}}
\end{figure*}

\begin{figure*}
\centering
\includegraphics[width=\textwidth]{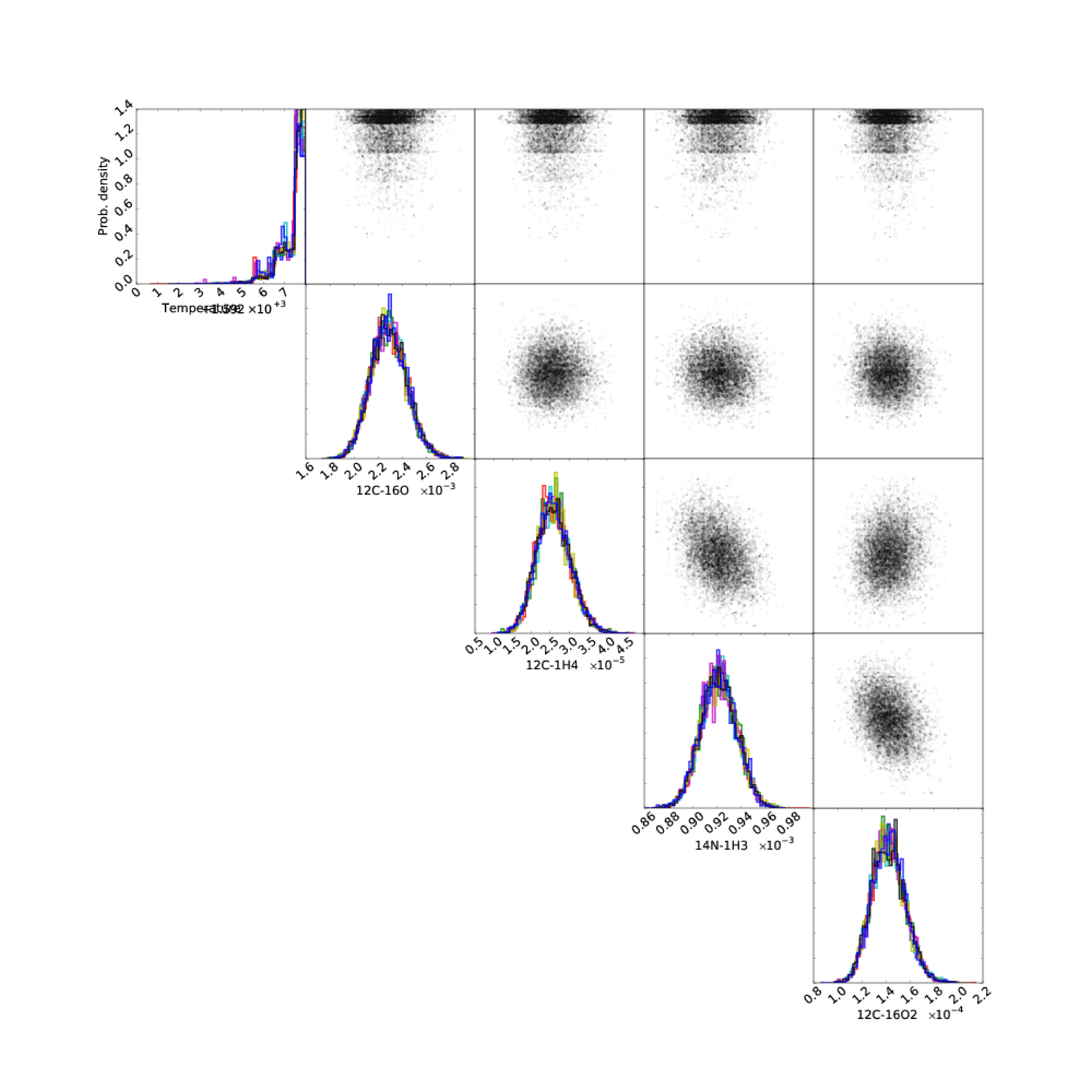}
\caption{showing marginalised and conditionals posterior distributions of the MCMC run for the under-complete model in section~\ref{sec:example-model}. Similar to figure~\ref{fig:post-nested-undercomplete} the MCMC is converging to the upper temperature prior rapidly.  This rapid convergence leaves `step' artefacts in the temperature posterior due to the discrete temperature resolution of 1K. Step sizes can be set to an arbitrarily small value but in this case convergence to the upper prior bound is fast enough to only ever sample the top bins after the burn in period is completed.} Otherwise identical to figure~\ref{fig:post-mcmc-complete}. \label{fig:post-mcmc-undercomplete}
\end{figure*}

\subsection{Model Selection}
\label{sec:example-model}

\begin{table}
{\small
\caption{Savage-Dickey density ratio (SDR) for overcomplete model } \label{table:sdrexample}
\begin{tabular}{lrrrrrr}
\hline\hline
   2ln($SDR$)    & TiO       & CO           &CH$_{4}$          & H$_{2}$O           &NH$_{3}$         &CO$_{2}$ \\\hline
     NS                  & 5.5       & -31.9        & -0.5                    & -31.9                    & 1.9                     & -31.9 \\
MCMC                & 10.3     & -29.3        & 1.6                     & -29.3                    & 3.6                     & -29.3 \\
\hline 
\end{tabular}
 }
\end{table}

In addition to the complete model shown above, we have simulated an over-complete and under-complete model to test the model selection abilities of \taurex. The over-complete model contains TiO as additional absorber and the under-complete model lacks H$_{2}$O. Whereas in terms of $\chi^{2}$ statistics we would expect a similarly valid fit for the over-complete model compared to the complete case (as the excessive parameters should converge to small values), we would expect a decrease in the overall model evidence as well as a clear discrimination of unnecessary complexity in the SDR. This behaviour is indeed demonstrated by \taurex. Figure~\ref{fig:modelselection} (middle) shows the model fit of the over-complete model and figures~\ref{fig:post-nested-overcomplete} and \ref{fig:post-mcmc-overcomplete} the posterior distributions of the NS and MCMC fits respectively. As figure~\ref{fig:modelselection} shows, the fit is maintained but at a lower global evidence, log(E) = -980 compared to log(E) = -960 for the correct model. On the Jeffrey's scale this results in a very strong preference for the overall simpler (i.e. complete) model. Table~\ref{table:sdrexample} shows the SDRs calculated from the NS and MCMC posteriors for $ \mathcal{M}_{{\bm \theta}-\theta_{\gamma}}/ \mathcal{M}_{{\bm \theta}}$, where $\theta_{\gamma}$ is the molecule in question. The SDRs show a substantial to strong preference for the exclusion of TiO from the model and strongly confirm the inclusion of CO, H$_{2}$O and CO$_{2}$. For the two low column density species, CH$_{4}$ and NH$_{3}$, the SDRs neither include nor exclude either species but do not support a significant detection of the molecule in the data, as expected. Differences in the SDR derived between NS and MCMC are due to the NS providing a tighter constraint on the marginalised posterior distributions than the MCMC, see figures~\ref{fig:post-nested-overcomplete} \& \ref{fig:post-mcmc-overcomplete}.

Figure~\ref{fig:modelselection} (bottom) shows the under-complete model excluding water. For under-complete models the $\chi^{2}$ increases significantly as well as a very low global evidence of log(E) = -9232. Figures~\ref{fig:post-nested-undercomplete} and \ref{fig:post-mcmc-undercomplete} show the posterior distributions of the NS and MCMC runs respectively. Both show a strong over dependence on high atmospheric temperatures, trying to fill in the missing opacities with and increased absorption due to an increased planetary scale height and an increased spectral broadening through the emergence of molecular hot-bands at higher temperatures.

\subsection{Resolution and Signal-to-Noise}
\label{sec:example-res}

\begin{table}
\caption{Savage-Dickey Ratios (SDRs) for H$_{2}$O, CO and NH$_{3}$ for data error-bars of $\sigma =$ 10ppm, 50ppm, 100ppm, 500ppm and resolutions of R = 300, 200, 100, 50, 30. Negative values signify a detection of the molecule with values $<-10$ being a very strong detection. Similarly, positive values $>6$ strongly indicate a non-detection. Values between -2 and +2 are inconclusive. \label{table:sdrres}}
\begin{tabular}{r|rrrr}
\hline\hline
   Resolution &   10ppm &   50ppm &   100ppm &   500ppm \\
\hline
%H$_{2}$O &            &            &            &         \\
H$_{2}$O          300 &  -28.3 &  -28.3 &  -28.5 &   -1.6 \\
          200 &  -28.3 &  -28.6 &  -28.5 &   -2.6 \\
          100 &  -28.3 &  -28.4 &  -28.5 &   -1.9 \\
           50 &  -28.2 &  -28.2 &  -28.6 &   -0.8 \\
           30 &  -28.7 &  -28.7 &  -28.7 &   -1.2 \\
\hline
%CO           &            &            &            &         \\
CO          300 &  -28.3 &  -28.3 &  -28.5 &   -1.4 \\
          200 &  -28.3 &   -28.6 &  -28.5 &   -1.0   \\
          100 &  -28.3 &   -28.4 &   -2.7 &   -0.1 \\
           50 &  -28.2 &  -28.2 &   -2.3 &   -0.1 \\
           30 &  -28.7 &  -28.7 &   -2.1 &    0.2 \\
\hline
%NH$_{3}$ &            &            &            &         \\
NH$_{3}$           300 &    8 .0  &    5.7 &    4.8 &    9.1 \\
          200 &    6.6 &    2.3 &    4.4 &    7.9 \\
          100 &    0.5 &    3.0   &    3.6 &    6.6 \\
           50 &   -1.6 &    4.8 &    6.2 &    5.6 \\
           30 &    7.7 &    7.7 &    8.4 &    5.5 \\
\hline
\end{tabular}

\end{table}

\begin{figure*}
\centering
\includegraphics[width=\textwidth]{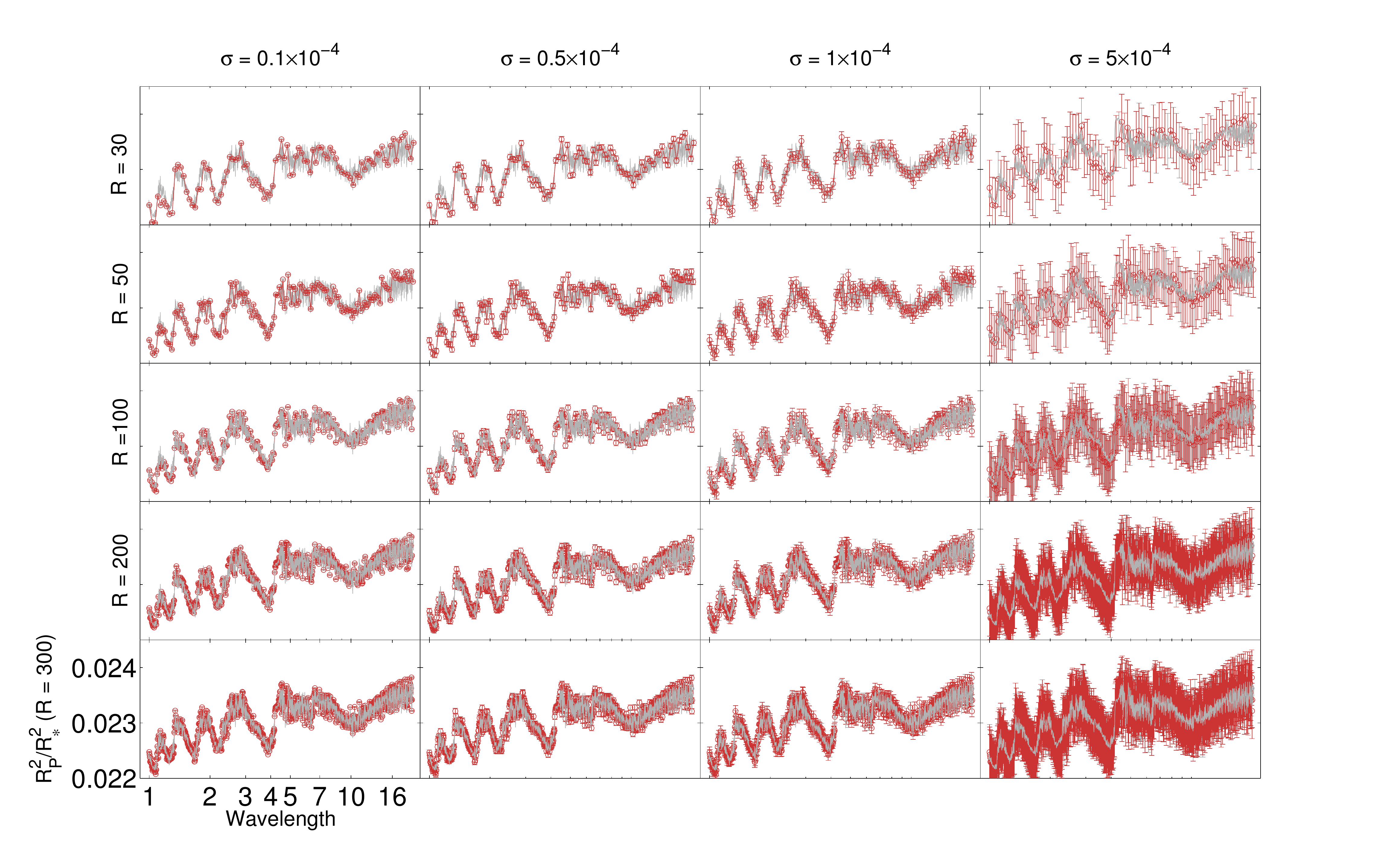}
\caption{Best fitting transmission model (grey) at R = 500 superimposed on simulated input data (red) at resolutions R = 30, 50, 100, 200, 300 and data-error bars $\sigma$ = 10ppm, 50ppm, 100ppm, 500ppm.    \label{fig:sens-spectra}}
\end{figure*}

\begin{figure}
\centering
\includegraphics[width=\columnwidth]{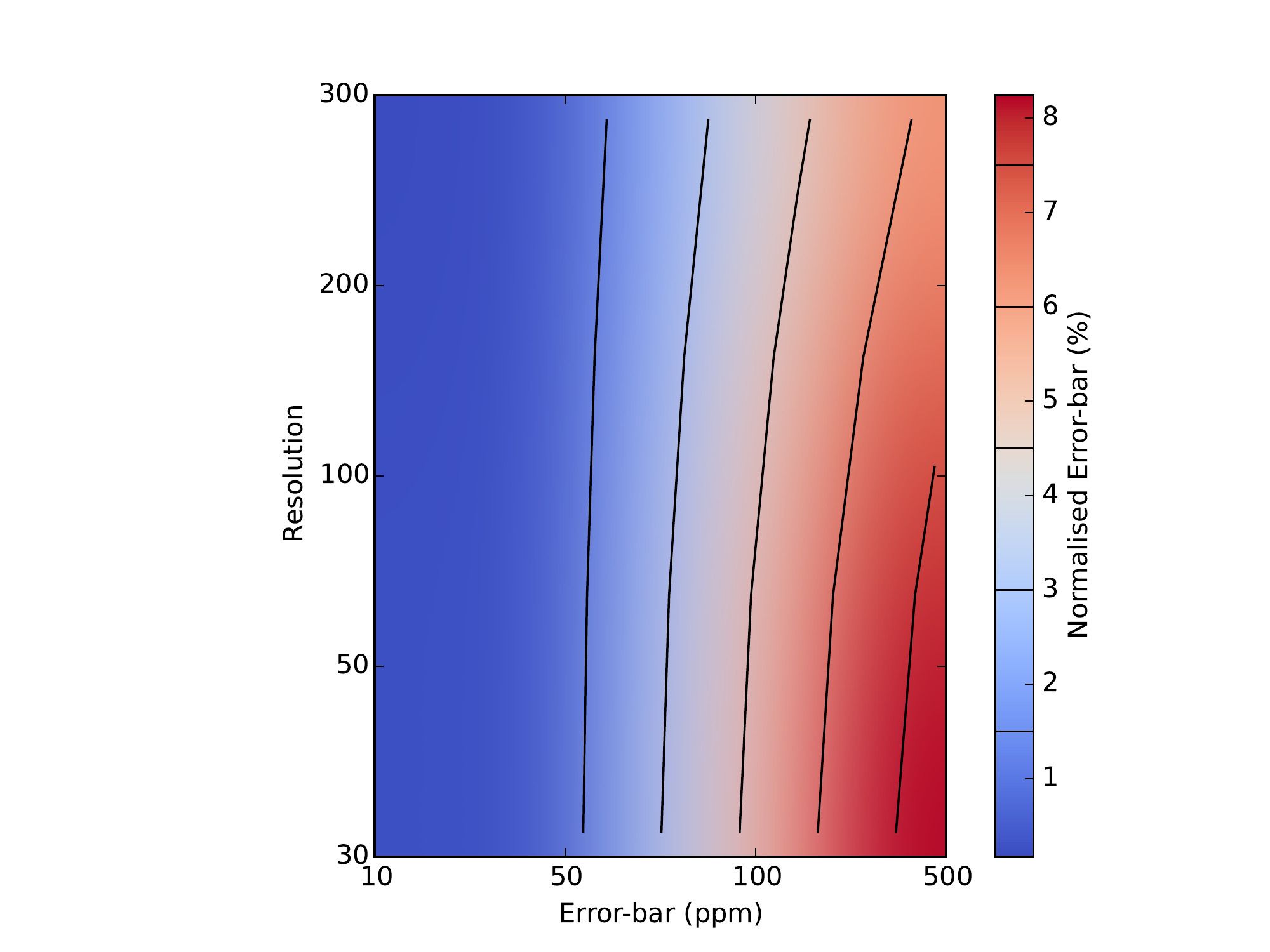}
\caption{Temperature posterior standard deviation normalised by the ground-truth temperature (1400K) as function of spectral resolution (R) and the data-error bar.   We find that the retrieval of the planetary temperature is dominated by the signal-to-noise of the data and less dominated by the resolution of the spectrum. \label{fig:sens-temp}}
\end{figure}

\begin{figure}
\centering
\includegraphics[width=\columnwidth]{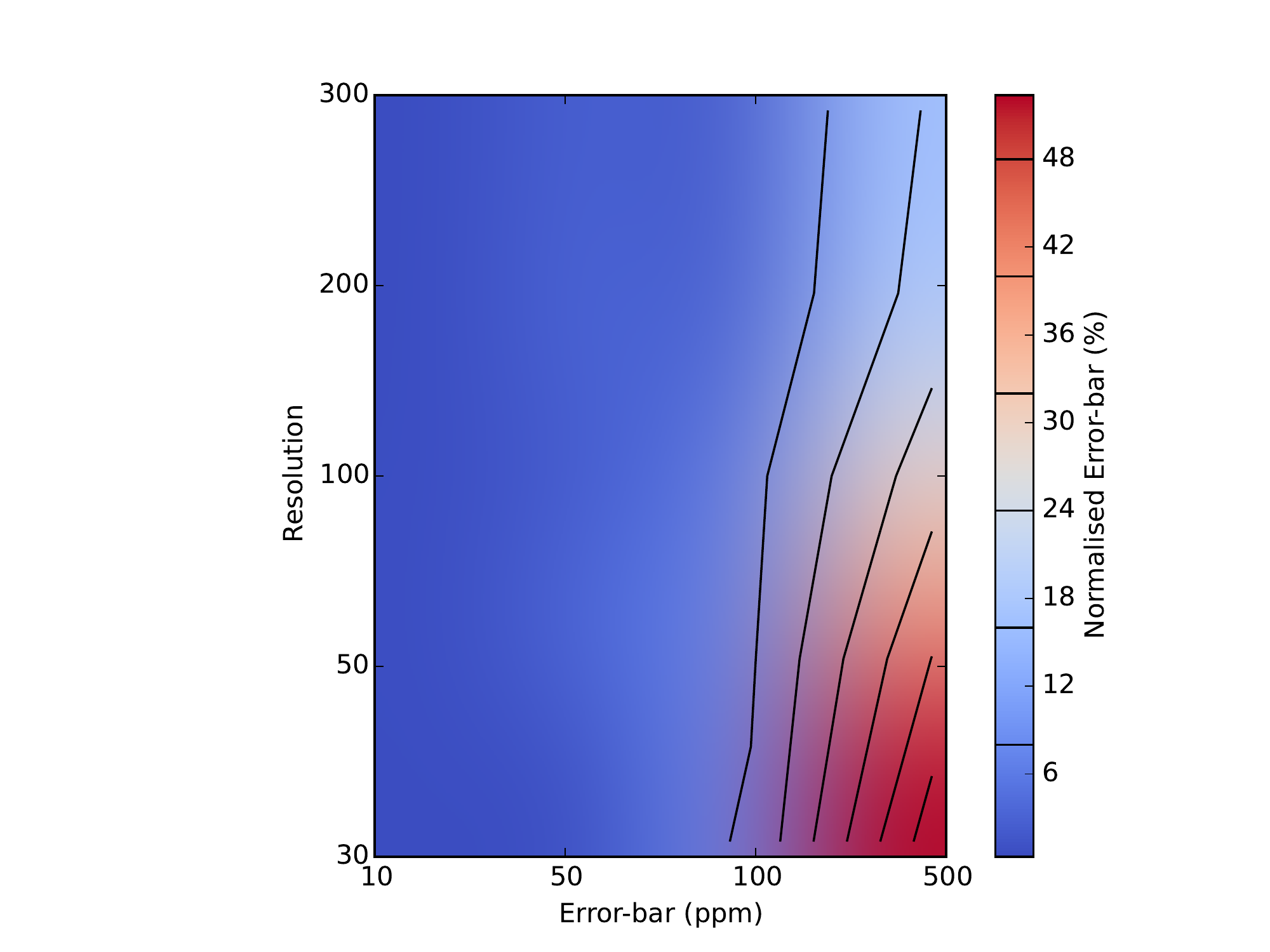}
\caption{H$_{2}$O posterior standard deviation normalised by the ground-truth abundance ($\chi_{(H_{2}O)}$ = 2$\times$10$^{-3}$) as function of spectral resolution (R) and the data-error bar.  The ability to retrieve water abundances remain relatively stable for high-resolution and low S/N data but significantly decreases as both S/N and R drop. Here posterior error-bars can reach the size of prior space.  \label{fig:sens-h2o}}
\end{figure}

\begin{figure}
\centering
\includegraphics[width=\columnwidth]{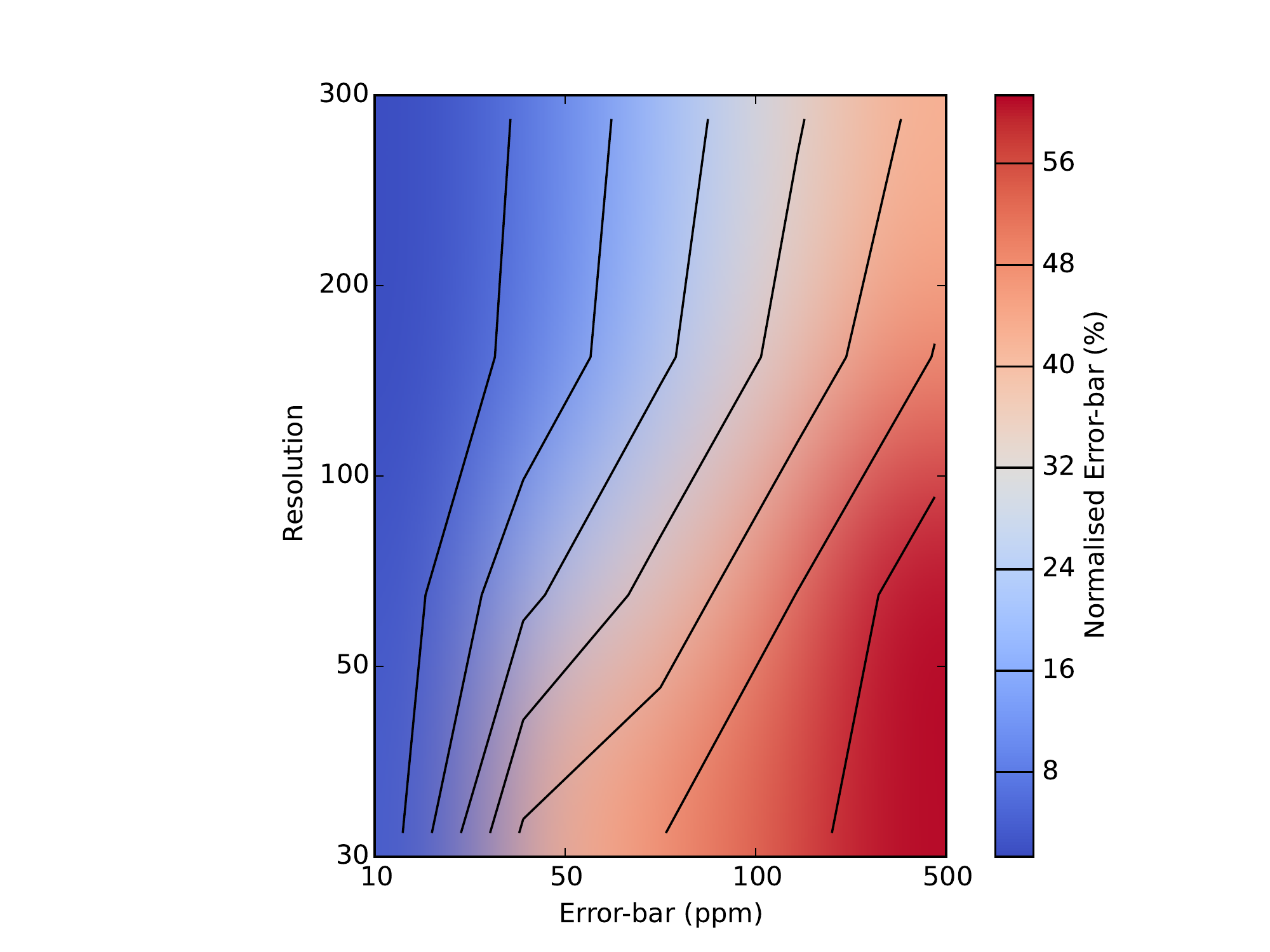}
\caption{CO posterior standard deviation normalised by the ground-truth abundance ($\chi_{(CO)}$ = 2$\times$10$^{-3}$) as function of spectral resolution (R) and the data-error bar.  The CO retrieval more strongly depends on S/N other than for the resolution grid considered here.\label{fig:sens-co}}
\end{figure}

\begin{figure*}
\centering
\includegraphics[width=\textwidth]{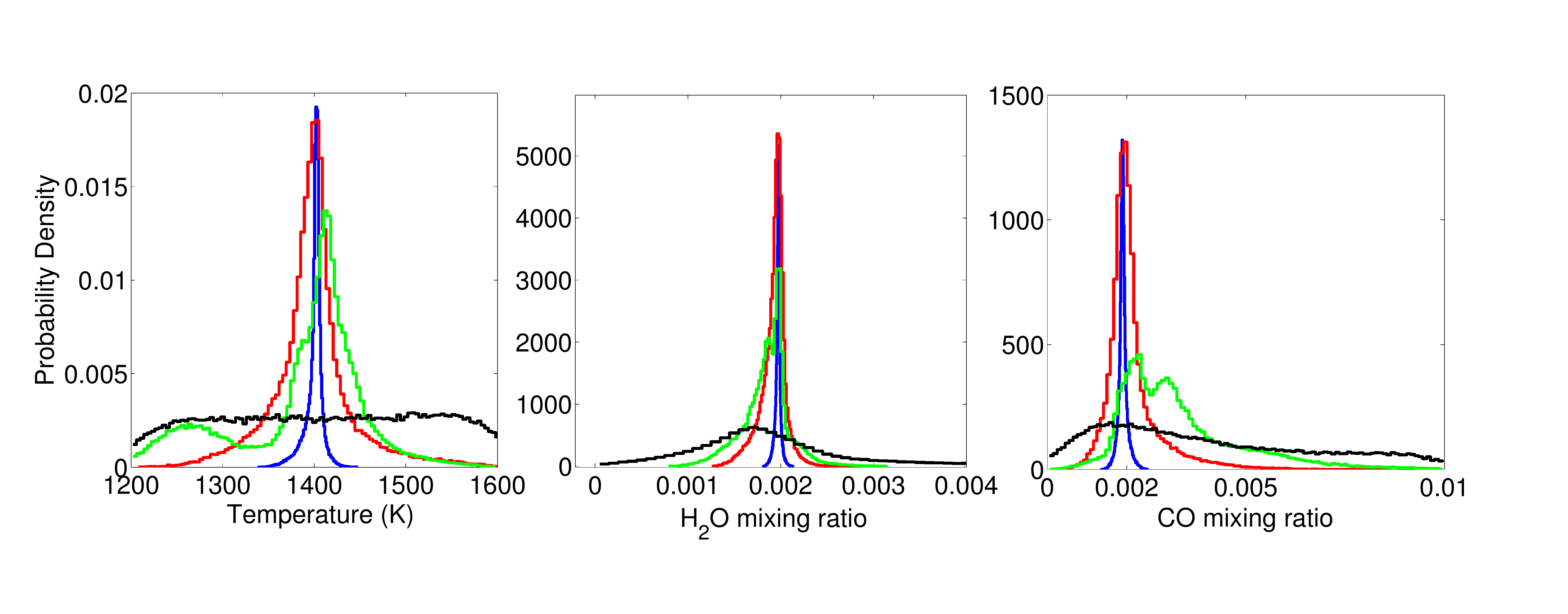}
\caption{Three figures showing the posterior distributions for the planetary temperature (left), H$_{2}$O (middle) and CO (right) mixing ratios at resolution R = 200. Colours represent data error bars. Blue: $\sigma =$ 10ppm (scaled by a factor or 1/5 to improve comparability to other curves); Red: 50ppm; Green: 100ppm; Black: 500ppm.   \label{fig:sens-r200}}
\end{figure*}

We now take the complete model from the previous section and reduce the signal to noise (S/N) and resolution to explore the impact on the retrieval of exoplanetary spectra. Given the potentially large scope of such an exercise we here limit ourselves to a S/N-Resolution grid most representative of current data: R = 300, 200, 100, 50, 30 and spectral error bars of $\sigma =$ 10ppm, 50ppm, 100ppm and 500ppm. Figure~\ref{fig:sens-spectra} shows the input data in red and the best fitting transmission model at a resolution of R = 500. Whereas visually all spectra fit equally well, degeneracies between mixing ratios and temperature increase as resolution and S/N decrease. Whereas this result is intuitive, we find that the effect of degrading resolution and S/N is not uniform amongst parameters. Figure~\ref{fig:sens-temp} shows the increase in error-bar (in percent) for the temperature posterior distribution derived. Here the reduction in S/N (i.e. increase in $\sigma$) has a much more pronounced effect than the reduction in resolution, meaning that the planetary temperature can be derived with high confidence for low resolution data but not vice versa. Figure~\ref{fig:sens-h2o} shows the same plot for the retrieval of the water mixing ratio. Water being very broad absorber in the NIR to mid-IR its abundance retrieval depends on resolution and S/N in approximately equal measures. Figure~\ref{fig:sens-co} plots the retrieval error bars of carbon-monoxide. With the spectral feature of CO being less broad than water we see that the dependence on S/N exceeds that of resolution (assuming that each CO feature is captures by at least one data point).  

The loss of information through a reduced S/N is also demonstrated in figure~\ref{fig:sens-r200} showing the posterior distribution for all $\sigma$ considered at R = 200. Here we can see the transition of the Bayesian argument going from data to prior dominated as the S/N decreases. Such a transition is gradual and we find a more regular occurrence of small local likelihood maxima as the likelihood surface ``flattens out'' at increasing $\sigma$. This is demonstrated by the the local maxima in the green curve ($\sigma =$ 100ppm) compared to the best determined solution (blue curve, $\sigma =$ 10ppm) and the prior driven solution (black curve, $\sigma =$ 500ppm). 
The theoretical behaviour of a Bayesian retrieval at low R and low S/N is an important result which will be address in detail in future work. 

Finally, we explore the effect of S/N and R on model selection. Similar to table~\ref{table:sdrexample} we calculated the SDR for H$_{2}$O, CO and NH$_{3}$ for the above S/N and R grid (table~\ref{table:sdrres}). As with previous examples negative values below $<-6$ indicate a strong detection with $<-10$ being a decisive detection whilst equally positive values rule out a detection in the data. As expected, we see the detection evidence for H$_{2}$O and CO decrease at high $\sigma$ and low R. NH$_{3}$ (and CH$_{4}$ not shown) remain undetected throughout as expected.

\citet{2012ApJ...749...93L} present a complementary analysis for the emission spectroscopy case where the overall informational content, and the resulting possible number of retrievable free parameters, is calculated. Their analysis points at the inverse relationship between S/N and R. The lower the S/N of an observation, the higher the resolution must be to obtain the same degree of retrievability, and vice versa. This relationship we also find for the retrievability of individual parameters in the transmission case, e.g. CO in figure~\ref{fig:sens-co}.

\section{Summary \& Conclusion}

In this publication we have introduced the \taurex~retrieval code for exoplanetary atmospheres. As described in the introduction and shown throughout the text, we have based the design of \taurex~on three guiding principles:  1) Sensitivity, 2) Objectivity, and 3) Big data. 

\taurex~incorporates a line-by-line radiative transfer code using state-of-the-art molecular opacity line lists by the \exomol~project. Atmospheric transmission models are run at $\sim$50-100 times higher resolution than the observed data to ensure a correct treatment of optically thick absorption lines as well as allowing for a precise treatment of thermal line broadening through an arbitrarily finely sampled temperature grid. 

Given the large number of potential absorbing/emitting species of an extrasolar planet, we have developed custom build pattern recognition software (the \marple~module) to rapidly scan large molecular and atomic line-list archives for possible absorbing/emitting signatures in the observed spectrum. By not manually specifying a list of molecules `expected' to be present in the atmospheres of exoplanets we break potential human biases in the selection of the atmospheric model. In other words by not assuming anything about the atmospheres composition and structure we maximise the objectivity of the analysis from the start.

Whereas the \marple~module sets the potential prior space of the fully Bayesian retrieval, the \occam~module performs iterative Bayesian model selection and iteratively verifies the adequacy of individual parameters as well as the overall evidence of the atmospheric model itself. 

By using efficient MCMC and Nested Sampling techniques throughout, we are able to parallelise the sampling of the likelihood space making \taurex~natively scaleable to cluster computing. This allows \taurex~to explore very large parameter spaces and accurately map correlation manifolds. 

We demonstrated individual properties of \taurex~using a simulated hot-Jupiter and explored the model selection process of over-complete and under-complete models. The quality of the retrieval was investigated for varying resolutions and signal-to-noise ratios of the input data and found to be consistent with expectations. 

Future work will see a detailed treatment of emission spectroscopy in the framework of \taurex, explore modelling degeneracies over large and short wavelength ranges and see the application to individual data sets. 

With the maturation of data reduction techniques for exoplanetary spectroscopy we obtain higher and higher precision spectroscopy of these exotic atmospheres. With higher precision of the data often comes higher complexity in the interpretation. The goal of an ideal retrieval of atmospheric properties is to be able to capture said complexity whilst maintaining the highest possible degree of objectivity in the analysis. 
 \taurex~presents a significant step towards this goal.  

 \section*{Acknowledgements}
We thank the referee for providing useful comments. This work was supported by the ERC project numbers 617119 (ExoLights) and
267219 (ExoMol).

%The research leading to these results has received funding from the European Research Council under the European Union's Seventh Framework Programme (FP/2007-2013) / ERC Grant Agreement n. [xxxxxx] and [xxxxxx] .

\appendix
\section{DRAM}
\label{appendix:dram}

In standard Metropolis-Hastings samplers each proposal step can either be accepted or rejected based on a fitness criterion and often a probability of acceptance when the fitness criterion is not met. Should the proposal be rejected the MCMC chain remains in the same position on the likelihood space. The delayed rejection (DR) mechanisms allows for a second (and third) proposal attempt to be made which is dependent on the previous chain as well as the previously rejected proposals. The adaptive proposal distribution based on its past history furthermore increases the efficiency and accuracy of the chain's exploration as the proposal distribution is iteratively adapted to the target distribution. These features can be shown to significantly improve the efficiency of the MCMC chain in high dimensional likelihood space.

\section{Glossary}
\label{appendix:glossary}
%\begin{table}
{\small
\begin{tabular}{l | l | l}
Variable & Description & Equation example \\\hline
$ N$ & Number of spectral points in data & \\\hline
$N_{m}$ & Number of molecules selected for retrieval & \\\hline
$m$ & Molecular species index & \\\hline
$\lambda$ & Wavelength index & \\\hline
${\bf x}$ & Data column vector &  \\\hline
$\bar{\bf x}$ & Normalised data column vector &\ref{equ:norm-data}\\\hline
$\varsigma$ & Absorption cross section & \ref{equ:crosssec-interp} \\\hline
$T$ & Temperature (K) & \\\hline
$a, b$ & Absorption cross section temperature & \\ &interpolation coefficients & \ref{equ:crosssec-interp}, \ref{equ:crosssec-interp2}, \ref{equ:crosssec-interp3} \\\hline
$\chi$ & Atmospheric mixing ratio & \\\hline
${\bm \tau}$& Optical depth column vector over $\lambda$ & \\\hline
${\bf I}(z)$ & Intensity column vector over $\lambda$ as function of $z$ & \ref{equ:intensity} \\\hline
$z$ & Height in atmosphere &\\\hline
${\bm \alpha}$ & Total atmospheric absorption column vector & \ref{equ:atmabsorb} \\\hline
${\bm \alpha}_{m}(T,\chi)$ & Total atmospheric absorption column vector as function of &\\&molecule, temperature,mixing ratio& \ref{equ:pre-tau} \\\hline
$R_{p}$ & Planetary radius & \\\hline
$R_{\ast}$ & Stellar radius & \\\hline
${\bm \delta}$ & Transit-depth column vector & \ref{equ:transdepth} \\\hline
$\bf U$ & Left unitary matrix of single value decomposition &\ref{equ:pre-tau}, \ref{equ:pre-pca} \\\hline
$\bf \Sigma$ & Diagonal matrix of single value decomposition &\ref{equ:pre-tau}, \ref{equ:pre-pca} \\\hline
$\bf V$ & Right unitary matrix of single value decomposition &\ref{equ:pre-tau}, \ref{equ:pre-pca} \\\hline
${\bf pc}$ & Principal component vector & \ref{equ:pre-pca} \\\hline
$n$ & Principal component index & \\\hline
$\bm{\psi}_{m}$ & Boolean data masking vector as function of molecule and wavelength & \ref{equ:pre-boolean} \\\hline
$\eta$ & Molecule detection threshold coefficient & \ref{equ:pre-boolean} \\\hline
$\hat{\bf x}$ & Masked data vector & \ref{equ:pre-masked} \\\hline
$\widehat{\bf pc}$ & Masked PCA vector & \ref{equ:pre-pcamasked} \\\hline
$\mathfrak{d}_{m}$ & $l_{2}$-norm between normalised data  and 2$^{nd}$ principal component & \ref{equ:pre-l2}\\\hline
$f(\mathfrak{d}_{m})$ & Monotonically increasing function of $\mathfrak{d}_{m}$ & \ref{equ:pre-fd} \\\hline
$\phi$ & \marple~ cluster index & \ref{equ:pre-fd} \\\hline
$\varphi$ & \marple~ cluster index for highest second derivative of $f(\mathfrak{d})$ & \ref{equ:pre-cluster} \\\hline
$m_{select}$ & \marple~ determined molecular/atmoic species & \\\hline
$\mathcal{M}$ & Exoplanet model & \\\hline
$\theta_{\gamma}$ & Generic parameter of model $\mathcal{M}$ & \\\hline
${\bm \theta}$ & Column vector of parameters of model $\mathcal{M}$ & \\\hline
$\sigma_{\lambda}$ & One sigma error at wavelength $\lambda$ & \\\hline
$P(\theta | {\bf x}, \mathcal{M})$ & Posterior probability distribution of $\theta$ given ${\bf x}$ and $\mathcal{M}$ & \ref{equ:bayesequation} \\\hline
$P({\bf x} | \theta, \mathcal{M})$ & Likelihood distribution of $\theta$ & \ref{equ:bayesequation} \\\hline
$P(\theta, \mathcal{M})$ & Prior distribution of $\theta$ & \ref{equ:bayesequation} \\\hline
$P({\bf x} | \mathcal{M})$& Bayesian partition function & \ref{equ:bayesequation} \\\hline
$E$ & Bayesian Evidence & \ref{equ:bayesevidence} \\\hline
\end{tabular}
}
%\end{table}

\bibliographystyle{apj}
\bibliography{apj-jour,taurex-lib_revised}

\end{document}